\documentclass[%
 reprint,superscriptaddress,
 amsmath,amssymb,
 aps,
]{revtex4-2}
\usepackage{natbib}
\usepackage[normalem]{ulem}

\usepackage{threeparttable}
\usepackage{graphicx}
\usepackage{dcolumn}
\usepackage{bm}
\usepackage{lipsum}
\usepackage{color}

\begin{document}


\title{Barocaloric effect of elastomers and plastic crystals: an outlook}

\author{V. M. Andrade}\email{vandrade@ifi.unicamp.br}
\affiliation{IFIMUP - Institute of Physics for Advanced Materials, Nanotechnology and Photonics of University of Porto, Portugal}
\affiliation{Faculty of Engineering of University of Porto, Portugal}
\author{L. S. {Paix\~ao}}\email{lpaixao@ifi.unicamp.br}
\affiliation{`Gleb Wataghin' Institute of Physics, State University of Campinas (UNICAMP), Campinas-SP, Brazil}
\author{A. M. G. Carvalho}\email{amgcarvalho@unifesp.br}
\affiliation{Mechanical Engineering Department, Maring\'a State University, 87020-900, Maring\'a, PR, Brazil}
\affiliation{Chemical Engineering Department, S\~ao Paulo Federal University, 09913-030, Diadema, SP, Brazil}
\author{V. Franco}\email{vfranco@us.es}
\affiliation{Dpto. Física de la Materia Condensada. Universidad de Sevilla. P.O. Box 1065. 41080-Sevilla, Spain}
\author{M. S. Reis} \email{marior@id.uff.br}
\affiliation{Dpto. Física de la Materia Condensada. Universidad de Sevilla. P.O. Box 1065. 41080-Sevilla, Spain}
\affiliation{Institute of Physics, Fluminense Federal University, Av. Gal. Milton Tavares de Souza s/n, Niteroi-RJ, Brazil}

\date{\today}

\begin{abstract}
The urge for efficient and environmentally friendly alternatives for the current gas-based refrigeration is becoming more critical due to global warming and overpopulation. Among the main candidates, solid-state-based cooling technology is appealing to substitute the conventional systems. Recently, attention has been turned to the barocaloric effect in advanced materials, namely, plastic crystals and elastomers, with reports revealing great caloric responses around room temperature for a series of materials. In this sense, this review has the goal of collecting, organizing and comparing the results from the literature to provide an outlook on the subject.
\end{abstract}

\maketitle


\section{Introduction}

Conventional gas compression/decompression refrigeration relies on harmful substances with high global warming potential. In view of the huge environmental challenges humankind faces nowadays, eco-friendly technologies are of utter importance. Another disadvantage of conventional refrigeration is its difficulty to scaling down to small sizes aiming its use for electronic components. In this scenario, solid-state-based cooling is considered a viable alternative refrigeration technology, mainly for its environmental friendliness appeal.

Solid-state cooling is based on caloric effects, which are thermal responses of materials induced by external stimuli. Depending on the applied field, the effect is named: magnetocaloric, when induced by magnetic fields; electrocaloric, by applying electric fields; and mechanocaloric, when the material is under mechanical stresses. The mechanocaloric effect encompasses the elastocaloric and the barocaloric effects, induced by uniaxial stress and hydrostatic pressure, respectively.

In the last decades, a number of reports appeared describing new materials with increasing caloric performances; however, further performance improvement is needed to make an economically viable alternative for household refrigeration \cite{ismail2021review}. In recent years, elastomers and plastic crystals are classes of materials that received considerable interest due to their impressive optical, mechanical, and barocaloric properties \cite{miliante2020unveiling,das2020harnessing,lloveras2019colossal,lloveras2021advances}. From the environmental point of view, besides contributing to the reduction of greenhouse gas emissions (characteristic of the solid-state technology), elastomers showed a recycling appeal, through the possibility of blending with rubber waste materials~\cite{bom2020tire}.

Therefore, the present review aims to summarize and compare recent results from the literature on the barocaloric properties of elastomers and plastic crystals. We review the appropriate experimental techniques and the results thereby obtained.

\section{Materials survey}


\subsection{Plastic Crystals (PC)}

\textit{Plastic crystals (PC)} are molecular materials with distinct and remarkable features, being in a mesophase between liquid crystal (LC) and ordered crystal (OC) \cite{silva2019ordered}. The main distinction of PCs over LC and OC is regarding the degree of freedom for the molecules inside the structures. For LC, there is no formation of lattices, allowing translational and rotational motion of molecules, which show weak van der Waals interactions and long-range orientational ordering \cite{zhu2020dissecting}. Distinctly, OC implies the formation of a lattice with no accessible rotation and translation modes for the molecules, presenting a long-range order and high inter- and intra-molecular interactions. In between, the PC molecules are orientationally disordered in a well-defined lattice, with low inter- and intra- molecular interactions of long-range kind, giving rise to the observation of high plasticity for these materials.

The classical PC is made of molecules with regular shapes (globular or cylindrical), i. e., symmetrical around their center of mass or by rotation around an axis. The exploitation of PC dates from the early 1960s with the studies of materials comprising simple neutral molecular systems like chloroform, cyclohexane, alkanes, cycloalkanes, certain terpenoids, pivalic acid, pentaerythritol and its derivatives \cite{das2020harnessing}. Further investigations on binary compositions of anionic and cationic species, the so-called ionic PC, revealed that cylindrical and disk shape molecules can also lead to the formation of such interesting systems \cite{zhu2019organic}. However, techniques to predict if a material will form a PC or not is still under debate within the community \cite{silva2019ordered,chandra1991transitions}. The development of novel PC materials and the understanding of their properties is essential for applications in photonics, optics, energy and memory devices, just to name a few \cite{das2020harnessing}.

J. Timmermans was the first to identify the peculiarities of PC during the preparation of a data survey for organic crystalline compounds, as described in his review written in 1961 \cite{timmermans1961plastic}. For a set of compositions of simple constituents, a low entropy of fusion/melting ($\Delta S_f$) below 20 J/kgK was observed for systems composed of globular molecules, like methane, carbon tetrachloride and cyclohexane. This was the first empirical criterion to consider a material as PC: low $\Delta S_f$ and globular molecular shape. Currently, due to the advances \textit{in} these systems, to be considered a PC, the material should present other features like plastic flow under stress, high self-diffusion, high compressibility and high vapour pressures close to their high melting temperature (above 350 K) \cite{das2020harnessing}. For this reason, more recent studies have demonstrated that a PC can be formed even exceeding the limit established by Timmerman for $\Delta S_f$, i.e., with non-globular molecular shape structures \cite{zhu2019organic}. For the case of cylindrical and disk-like molecular shapes, there is a reduced number of accessible rotation modes that depend on the size and noncovalent interactions inside the structure, leading to $\Delta S_f$ values of 70 J/kgK - broadening the possibility of a system to form a PC. Materials with constrained available modes, namely OC, present $\Delta S_f$ higher than 80 J/kgK, which is the upper limit to classify a molecular arrangement as PC. These reduced $\Delta S_f$ values for plastic crystals are due to the highest degrees of freedom accessed by the molecules, which is also the reason for the crystalline arrangement presented by these structures.

\begin{figure}
   \centering
    \includegraphics[width=\columnwidth,keepaspectratio]{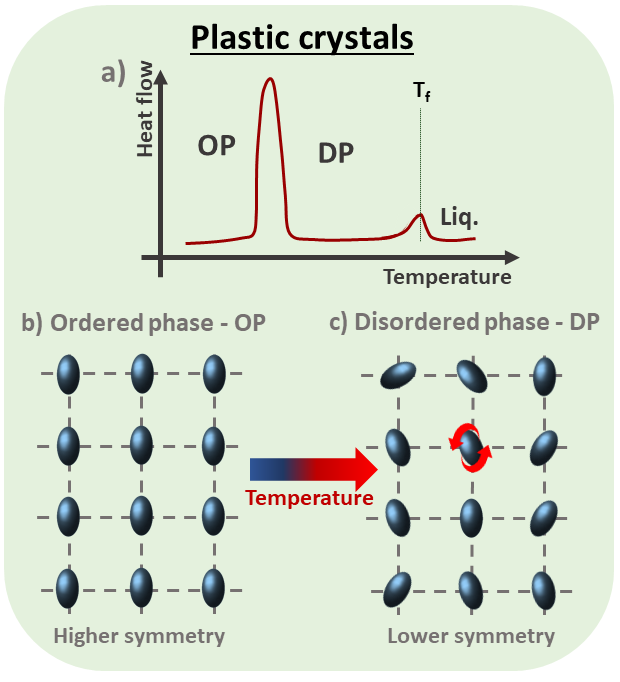}
   \caption{\textbf{(a)} An illustration of the temperature dependence of the heat flow for a plastic crystal, revealing that the contribution at the fusion temperature is lower than that at the order-disorder transition. Scheme of the plastic crystals \textbf{(b)} at the ordered phase at low temperature, where the molecules present a preferred orientation and \textbf{(a)} at the disordered phase with the reorientation modes available for the molecules above the transition temperature.}\label{fig:pc}
    \end{figure}

A general curve of heat flow \textit{vs.} temperature is given in Fig. \ref{fig:pc}(a), where the lower contribution, located at a higher temperature, arises from the melting process. 
Additionally to the properties mentioned above, the main characteristic of PCs is to present one or more solid-solid phase transitions from an ordered to a disordered state before melting, preferably reversible \cite{das2020harnessing}. Some PC systems can present successive order-to-disorder phase transitions, such as the trimorphic quinuclidium perrhenate that shifts from a cubic ordered phase to a pseudo-rhombohedral phase and then to an orthorhombic phase with lower symmetry \cite{harada2018ferroelectricity}. The system presents a higher symmetry state with preferred orientations for the molecules at low temperatures, as given in Fig. \ref{fig:pc}(b), usually hexagonal and cubic phases. By increasing the temperature, the possible rotational modes of the molecules can be accessed, and an anharmonic lattice deformation shifts the structure to a lower symmetry state, such as rhombohedral and monoclinic ones. During the symmetry change process, in a first-order transition, a large enthalpy change is observed due to the activation of the rotational modes by the molecules, as highlighted in figure \ref{fig:pc}(c). Although the plastic deformations are very isotropic, the molecular interactions can lead to irreversibility during the phase change during the application of stress \cite{aznar2020reversible}. Close to the fusion temperature (T$_f$), a high plasticity behavior and diffusion constant occur for PC due to the higher degrees of freedom and long-range orientational correlations for the molecules. As for the melting (fusion) process, lower enthalpy changes are associated, since the higher degree of freedom was already achieved by the system. In other words, the energy cost for the transition from an ordered to a disordered phase is higher than during the material melting, leading to greater entropy changes at the solid-solid phase transitions \cite{das2020harnessing}. This is one of the main reasons for the great potential of PCs as barocaloric materials, as will be further described in this review.  

\subsection{Elastomers}

Polymers have a variety of features and, consequently, these materials can be classified based on their source (natural/synthetic), structure, chemistry, connectivity (or tacticity) and spatial arrangements~\cite{sperling-book}. These features rank the polymers as elastomers, plastics and fibers. Plastics are materials with high molar weight and may be divided into \textit{thermoplastic} and \textit{thermosetting}. The former, having no cross-linked chains, is easily molded by heating, while the second presents hardening after heating due to the cross bond of the polymer chains. Fibers present higher structural organization, leading to less elasticity, higher intermolecular forces and a geometrical condition of length/diameter ratio higher than 100 \cite{sperling-book}. A scheme about the type
of polymers is given in Fig. 2, where these are aligned regarding to their mechanical properties (softness and hardness). Depending on the type of process to produce the polymers, they can present amorphous and crystalline phases, being, in reality, semicrystalline materials. Considering the Young's modulus dependence with temperature for an amorphous polymer, the system will present a rigid glassy state at low temperatures. As the temperature increases, a rubbery behavior occurs above the so-called glass transition temperature (T$_G$). The region of interest for the barocaloric effect relies on the rubbery state, i.e., the region between T$_G$ and the melting temperature (T$_M)$; which should be wide enough for practical use of the material. Above the glass transition~\cite{sperling-book}, their flexibility increases, reducing the Young's modulus. For this reason, considering flexibility as the requirement for applications, T$_G$ well below room temperature are preferable for amorphous elastomers. Among the mentioned polymer' classes, elastomers present higher elasticity and, for this reason, are the ones with higher barocaloric response, as will be further discussed.

\begin{figure*}
    \centering
    \includegraphics[width=15cm,keepaspectratio]{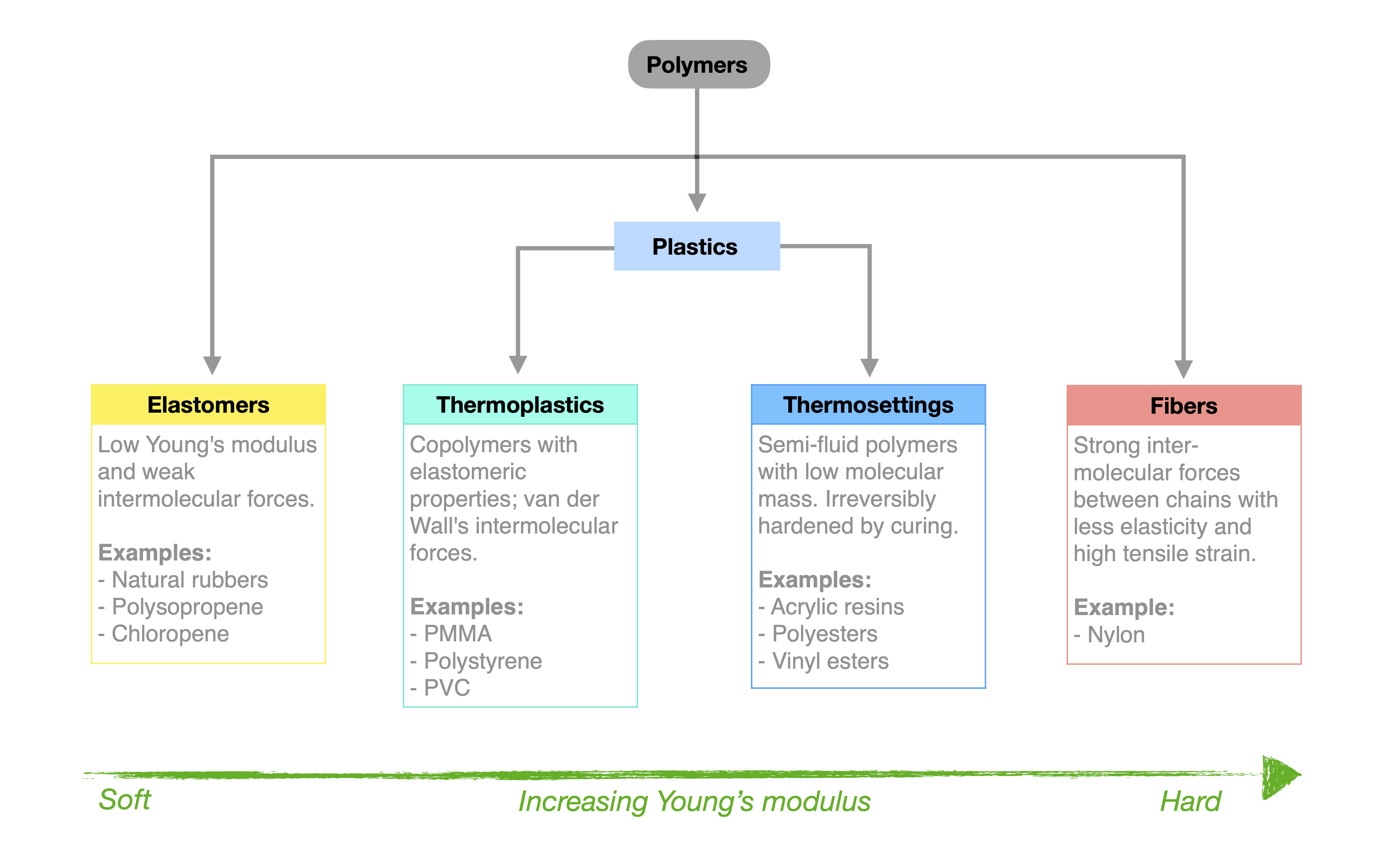}
   \caption{Classification of polymers according to their Young's modulus: from soft (left) to the hard (right).}
   \label{fig:polymer_class}
\end{figure*}

\textit{Elastomers} belong to the particular category of polymers that gave high failure strain, low Young's modulus and weak intermolecular forces - Coulomb and van der Waals interactions - being the main characteristics responsible for their rubber-like behavior. These materials can be classified as: \textit{natural}, when originated from plants and some fungi species; \textit{semi-synthetic}, when processed from other materials to improve resistance and elasticity~\cite{vijayaram2009technical}; and \textit{synthetic}, as petroleum or silicon-based products. In particular, 
natural rubber was used by ancient cultures of Mesoamerica as early as 1600 B.C.~\cite{hosler1988mesoamerica}; however, it only became popular after the development of the vulcanization process. These semi-synthetic polymers are commonly obtained by curing the sticky extracted rubber with sulfur, peroxides, bisphenol and carbon black to improve their mechanical strength, elasticity and durability~\cite{vijayaram2009technical}. Their large molecules may be cross-linked, as in the case of vulcanized natural rubber, or not, as in the case of raw (non vulcanized) natural rubber. Given this, elastomers became essential products for the modern world, as they are used in a vast range of industries due to their lightweight, flexibility, non-toxicity, low cost and sustainability, just to name a few advantages.

These materials present lower molecular weight and none or few cross-linked chains that are randomly oriented, having the freedom to rotate and translate in the structure. For this reason, the elastomers can be reversibly extended at room temperature up to twice their size without breaking~\cite{rosato2003plastics}, being of interest for mechanocaloric applications. 
Furthermore, it should be pointed out that at some conditions of temperature and pressure, the elastomers can rearrange their chains and, thus, present a crystalline structure~\cite{furushima2018crystallization}. Upon cooling, below the liquid flow region, the material presents a crystalline phase with lower symmetry and volume, being called the plastic phase. At the transition temperature, the system undergoes a first-order solid-state transition in which its molecules exhibit preferred orientations and, consequently, an ordered phase. The phase transition can be driven either by decreasing temperature, as mentioned above, or by increasing pressure, as it also limits the movement of the molecules and makes them ordered. Such behavior is observed through the appearance of characteristic peaks in X-ray diffraction measurements. This is the case of acetoxy silicone rubber (\textbf{ASR}), where a crystalline ordering occurs around 250 K, contributing to an extra increase in the barocaloric effect in this material\cite{imamurasupergiant}. However, it was recently shown through simulations that the rising of great BC response in elastomers originates at the molecular level. These aspects will be presented in more detail later in this review article.

\section{Theoretical basis}
\label{section_theory}

\subsection{Thermodynamics}

The first law of thermodynamics, written in terms of the strain tensor $\epsilon_{ik}$ (which is dimensionless) and the stress tensor $\sigma_{ik}$ (which has the dimension of pressure), reads as $
dE=TdS+V_0\sigma_{ik}d\epsilon_{ik}$, where the last term comprises a sum over the repeated indexes $i$ and $k$; and $V_0$ represents the volume of the unstressed material. From the Gibbs free energy $G=E-TS-V_0\sigma_{ik}\epsilon_{ik}$,
it is possible to write:
\begin{equation}\label{gibfnw}
   dG=-SdT-V_0\epsilon_{ik}d\sigma_{ik} 
\end{equation}
and, consequently, for an isothermal process ($dT=0$):
\begin{equation}\label{eq:strainstress}
    \left.\frac{\partial G}{\partial \sigma_{ik}}\right|_T=-V_0\epsilon_{ik}.
\end{equation}
In a similar fashion, for an isobaric process ($d\sigma_{ik}=0$), the relation 
\begin{equation}
 \left.\frac{\partial G}{\partial T}\right|_\sigma=-S
\end{equation}
holds and then it is possible to obtain an important Maxwell relation:
\begin{equation}\label{masiwe}
\left.\frac{\partial S}{\partial\sigma_{ik}}\right|_T=V_0\left.\frac{\partial\epsilon_{ik}}{\partial T}\right|_\sigma.
\end{equation}

The Helmholtz free energy $F=E-TS$ is also an important potential for the thermodynamic description of the caloric materials. Analogously to equation \ref{gibfnw}, it is possible to write:
\begin{equation}
   dF=-SdT+V_0\sigma_{ik}d\epsilon_{ik} 
\end{equation}
and, consequently, for an isothermal process, the following equation for the stress tensor holds:
\begin{equation}\label{eq:strainstress2}
    \left.\frac{\partial F}{\partial \epsilon_{ik}}\right|_T=V_0\sigma_{ik}.
\end{equation}

\subsection{Caloric potentials}

For the case of uniaxial stress, i.e., for elastocaloric studies,  the corresponding tensor is $\sigma_{ik}=\sigma\delta_{iv}\delta_{kv}$, where $v$ is the direction of the applied pressure. For this uniaxial case, $\sigma>0$ means tension and $\sigma<0$ means compression. On the other hand, $\sigma_{ij}=-p\delta_{ij}$ describes the barocaloric effect due to hydrostatic pressure. From these definitions, we are able to write different caloric potentials, focusing on the barocaloric effect.

\subsubsection{Isothermal entropy change}

Considering the barocaloric effect, equation \ref{masiwe} can be rewritten as:
\begin{equation}
-\frac{\partial S}{\partial p}=V_0\frac{\partial\gamma}{\partial T}
\end{equation}
where $\gamma=\epsilon_{ll}=\text{Tr}\{\epsilon_{ik}\}=\Delta V/V_0$. After integration, the isothermal entropy change, for a given pressure change, is given by:
\begin{equation}\label{DeltaSS}
    \Delta S(T,\Delta p)=-V_0\int_{p_1}^{p_2}\frac{\partial\gamma}{\partial T}dp.
\end{equation}

Around first-order transitions, a sharp change in the entropy takes place, often associated with a sharp change in the sample volume. This change across the transition ($\Delta S_t)$ is given by the Clausius-Clapeyron equation
\begin{equation}
\Delta S_{t}(T,\Delta p)=\Delta V_t\left(\frac{dT_t}{dp}\right)^{-1},
\end{equation}
where $\Delta V_t$ is the volume change of the material across the transition and $T_t$ is the transition temperature. This relation gives the entropy change induced by the transition. For materials presenting other relevant contributions to the entropy change, this approach does not retrieve the total entropy change. 

\subsubsection{Adiabatic temperature change}

In order to write the other caloric potential, namely, the adiabatic temperature change, we start considering the entropy as a function of temperature and stress tensor: $S=S(T,\sigma_{ik})$. Thus, it is possible to write:
\begin{equation}
    dS=\left.\frac{\partial S}{\partial T}\right|_{\sigma} dT+
    \left.\frac{\partial S}{\partial \sigma_{ik}}\right|_T d\sigma_{ik}.
\end{equation}
The adiabatic condition imposes $dS=0$ and therefore:
\begin{equation}
    dT=-V_0\frac{T}{c_\sigma}\left.\frac{\partial \epsilon_{ik}}{\partial T}\right|_{\sigma}d\sigma_{ik},
\end{equation}
where we have used equation \ref{masiwe} and the definition of specific heat as $c_\sigma= T\ \partial S/\partial T|_\sigma$. From the equation above, the adiabatic temperature change can be obtained for the barocaloric case:
\begin{equation}\label{deltaTT}
    \Delta T(T,\Delta p)=V_0\int_{p_1}^{p_2}\frac{T}{c_p}\left.\frac{\partial \gamma}{\partial T}\right|_{p}dp.
\end{equation}

An interesting and useful relationship emerges from equations \ref{DeltaSS} and \ref{deltaTT}. Considering the specific heat $c_p$ has a weak dependence with the applied pressure, it is possible to verify the following equation:
\begin{equation}\label{deltfwff}
    \Delta T(T,\Delta p)\approx-\frac{T}{c_p}\Delta S(T,\Delta p).
\end{equation}

\subsection{Landau approach}

\subsubsection{Equation of state}

The free energy can be expressed as a power function of the strain tensor (\textit{provided that its elements are small enough}):
\begin{align}\nonumber
F(T)=&F_0(T)-KV_0\alpha(T-T_0)\epsilon_{ll}\\\label{eq:f3ffwe}
&+\mu V_0\left(\epsilon_{ik}-\frac{1}{3}\delta_{ik}\epsilon_{ll}\right)^2+\frac{1}{2}KV_0\epsilon_{ll}^2,
\end{align}
where the second term is related to the thermal expansion, the third is the shear term and the last one is the Hook's term. $K$ is known as the bulk modulus, $\mu$ is the shear modulus, $V_0$ is the volume of the unstressed material, T$_0$ is a reference temperature where the sample experiences no thermal deformation and, finally, $\alpha$ is the thermal expansion coefficient.

From equations \ref{eq:strainstress2} and \ref{eq:f3ffwe}, a general expression for the stress tensor can be obtained:
\begin{align}
    V_0\sigma_{ik}=\frac{\partial F}{\partial \epsilon_{ik}}=&-KV_0\alpha(T-T_0)\delta_{ik}\\
    &+2\mu V_0\left(\epsilon_{ik}-\frac{1}{3}\delta_{ik}\epsilon_{ll}\right)+KV_0\epsilon_{ll}\delta_{ik}.
\end{align}
Considering only an isothermal process, in which the temperature of the material is kept constant, i.e., $T=T_0$, the first term above is null and therefore:
\begin{equation}\label{eq:sigmaikfrever}
    \sigma_{ik}=K\epsilon_{ll}\delta_{ik}+2\mu\left(\epsilon_{ik}-\frac{1}{3}\delta_{ik}\epsilon_{ll}\right).
\end{equation}
Conversely, the strain tensor can be written as:
\begin{equation}\label{eq:eq:uikfwe}
    \epsilon_{ik}=\frac{\sigma_{ll}\delta_{ik}}{9K}+\frac{1}{2\mu}\left(\sigma_{ik}-\frac{1}{3}\sigma_{ll}\delta_{ik}\right).
\end{equation}

Let us write these equations in terms of more convenient parameters, the Young modulus $Y$ and Poisson ratio $\nu$, which relate to the shear and bulk moduli as  
\begin{equation}
    \mu=\frac{Y}{2(1+\nu)}\;\;\;\text{and}\;\;\;K=\frac{Y}{3(1-2\nu)}.
\end{equation}

In terms of the Young modulus and Poisson's ratio, the general equations \ref{eq:sigmaikfrever} and \ref{eq:eq:uikfwe} can be rewritten as:
\begin{equation}\label{eq:sigmjwer}
    \sigma_{ik}=\frac{Y}{1+\nu}\left[\epsilon_{ik}+\frac{\nu}{(1-2\nu)} \epsilon_{ll}\delta_{ik}\right]
\end{equation}
and
\begin{equation}
    \epsilon_{ik}=\frac{1}{Y}\left[(1+\nu)\sigma_{ik}-\nu\sigma_{ll}\delta_{ik}\right].
    \label{eq:uik}
\end{equation}

\subsubsection{Isotermal entropy change}

Considering the free energy (equation \ref{eq:f3ffwe}), the entropy of the system can be derived as:
\begin{equation}
    S(T,p)=-\frac{\partial F}{\partial T}=S_0(T)+KV_0\alpha \epsilon_{ll}.
\end{equation}
As we are interested in an isothermal process ($T=T_0$), the entropy change due to a pressure change, can be written as:
\begin{align}\nonumber
    \Delta S(T,\Delta p)&=S(T,p)-S(T,p_0)\\\label{dsdwefwef}
    &=KV_0\alpha \epsilon_{ll},
\end{align}
where the final result depends on how the compression is achieved. In this way, we can differentiate two cases: \textit{uniaxial} vs. \textit{isotropic} compression.

\textit{Uniaxial compression}: This case corresponds to a rod compressed along the $z$-axis, whose $x$ and $y$ sides are fixed. For this case, all strain components are zero, except $\epsilon_{zz}$. From equation \ref{eq:sigmjwer}, the stress components of the tensor can be written as:
\begin{equation}
    \sigma_{zz}^{uni}=\frac{Y(1-\nu)}{(1+\nu)(1-2\nu)}\epsilon_{zz}^{uni}
    \label{sigmazz}
\end{equation}
and
\begin{equation}
    \sigma_{xx}^{uni}=\sigma_{yy}^{uni}=\frac{Y\nu}{(1+\nu)(1-2\nu)}\epsilon_{zz}^{uni}.
    \label{sigmaxx}
\end{equation}
Considering $\sigma_{zz}^{uni}=-p$, the non-null strain component can be written in terms of the applied pressure $p$:
\begin{equation}\label{unilwdfnuz}
    \epsilon_{zz}^{uni}=-\frac{(1+\nu)(1-2\nu)}{Y(1-\nu)}p
\end{equation}

If we place the above result into equation \ref{dsdwefwef}, the isothermal entropy change is obtained for this specific case:
\begin{equation}\label{mceuniw}
    \Delta S^{uni}(T,\Delta p)=-\frac{\alpha}{3}\frac{(1+\nu)}{(1-\nu)}p V_0.
\end{equation}

\textit{Isotropic compression}: This case is characterized by equal stresses along with the diagonal components of the tensor $\sigma_{xx}^{iso}=\sigma_{yy}^{iso}=\sigma_{zz}^{iso}=-p$, while the off-diagonal components are zero. From equation \ref{eq:uik} it is possible to write:
\begin{equation}
\epsilon_{xx}^{iso}=\epsilon_{yy}^{iso}=\epsilon_{zz}^{iso}=-\frac{p}{Y}(1-2\nu).
\label{uzziso}
\end{equation}
Analogously to before, moving the above result into equation \ref{dsdwefwef} the following expression emerges:
\begin{equation}
\Delta S^{iso}(T,\Delta p)=-\alpha p V_0.
\label{eq:deltaSbaro}
\end{equation}

\subsubsection{Uniaxial compression vs. isotropic compression}

Isotropic compression can also occur from the uniaxial compression discussed before. From equations \ref{sigmazz} and \ref{sigmaxx}, the following relation reads:
\begin{equation}
\sigma_{zz}^{uni}=\sigma_{xx}^{uni}\frac{1-\nu}{\nu},
\end{equation}
When the Poisson coefficient approaches 0.5, the equation above reads as $\sigma_{zz}^{uni}\to\sigma_{xx}^{uni}$; i.e., the uniaxial compression effectively works as an isotropic compression. We can also compare the strain tensor for uniaxial and isotropic compression from equations \ref{unilwdfnuz} and \ref{uzziso}, obtaining:
\begin{equation}\label{poiseq}
\epsilon_{zz}^{uni}=\epsilon_{zz}^{iso}\frac{1+\nu}{1-\nu}.
\end{equation}
Again, in the limit $\nu\to 0.5$, the equation above reads as $\epsilon_{zz}^{uni}=3\epsilon_{zz}^{iso}$. For the uniaxial compression, the sample  is constrained to deform in only one direction (which we assume as the $z$ direction). Thus, the deformation is 3 times larger than the case of isotropic compression, where the sample is free to deform in all 3 directions. 

In fact, for materials with Poisson coefficient close to 0.5, as is the case of elastomers and natural rubbers, the BC of an uniaxial compression (equation \ref{mceuniw}) recovers the BC of an isotropic compression (equation \ref{eq:deltaSbaro}). Thus, the isotropic compression and the unilateral compression (for materials with $\nu\sim 0.5$), described above, are basic configurations for measuring barocaloric effects.\\

\section{Experimental techniques for the barocaloric effect}

It has been used two experimental methods to obtain the caloric parameters in plastic crystals and polymers. The first one depends on indirect entropy change calculation using calorimetric data under applied pressure. From these results, equation \ref{deltfwff} has been used to estimate the adiabatic temperature change. The second procedure uses a direct approach to measure the adiabatic temperature change upon cooling/heating and load/unloading the sample. A strain-gauge coupled to the setup allows the collection of strain versus stress data, to indirectly calculate the entropy change through equation \ref{DeltaSS}. Each well established procedure and its advantages/disadvantages is described below.

\begin{figure}
    \centering
    \includegraphics[width=\columnwidth,keepaspectratio]{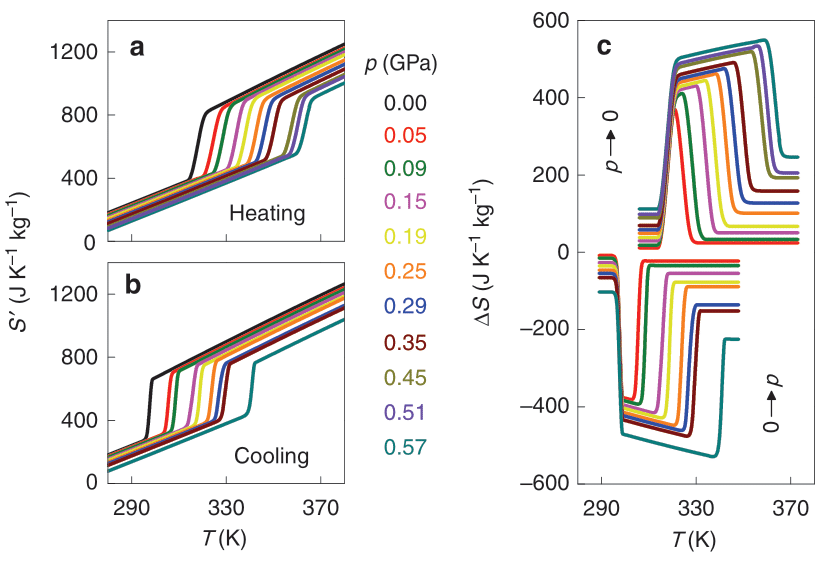}
    \caption{Entropy curves for neopentylglycol (\textbf{NPG}) calculated using equation \ref{S_PC} and calorimetric measurements on (a) heating/unloading and (b) cooling/loading. (c) The isothermal entropy change $\Delta S$ calculated from the loading/cooling curve in (b), with positive signal, and unloading/heating process in (a), with negative signal. This image is reprinted from reference \cite{lloveras2019colossal} with permission of Nature Publishing Group.}
    \label{fig:dS_signal}
\end{figure}

\textbf{Indirect measurements.} Obtaining caloric properties through heat capacity measurements is a well-established method for the community, being the first method commonly reported in the literature to estimate the barocaloric effect in different classes of materials. The procedure uses differential scanning calorimetry data and, in some reports, is complemented using temperature-dependent XRD data for the crystalline phases~\cite{lloveras2019colossal,aznar2020reversible}. It is claimed that the lattice parameter vs. temperature data is necessary to correct the isothermal entropy change obtained from calorimetric data~\cite{lloveras2019colossal}. In order to build the entropy change curves, a series of parameters are required: specific heat of ordered and disordered phases ($C_o$ and $C_d$, respectively), heat flow ($dQ/dT$) and the entropy change due to the volumetric expansion $\Delta S_{vol}(p) = -[(\partial V/\partial T)_{p_o}]p$, with $(\partial V/\partial T)_{p_o}$ considered as pressure independent - note that $p_o$ corresponds to the ambient pressure. The total $S'(T,p) = S(T,p) - S(T_o,p_o)$ equation, where $T_o$ is the reference temperature, reads then as~\cite{lloveras2019colossal}:
\begin{widetext}
\begin{equation}
S'(T,p)=\left\{
\begin{array}{ll}
\int_{T_o}^{T} \frac{C_o(T')}{T'}dT' + \Delta S_{vol}(p) & T<T_i\\
S(T_i,p) + \int_{T_i}^{T}\left[C_{o-d}(T') + \left|\frac{dQ(T',p)}{dT'}\right|\right]dT' + \Delta S_{vol}(p) & T_i<T<T_e\\
S(T_e,p) + \int_{T_e}^{T}\frac{C_d(T')}{T'}dT' + \Delta S_{vol}(p) & T>T_e
\end{array}\label{S_PC}
\right.
\end{equation}
\end{widetext}
where $C_{o-d} = (1-x)C_o + xC_d$ corresponds to the intermediate specific heat regarding to the disordered phase fractions ($x$) obtained from XRD analysis, $T_i$ and $T_e$ are the temperatures corresponding to the start and end of the ordered to disordered transformation, respectively, and $S(T_i,p)$ and $S(T_e,p)$ the corresponding values of entropy at those specific temperatures. An example of the constructed $S'(T,p)$ for neopentylglycol (\textbf{NPG}) is given in Fig. \ref{fig:dS_signal}(a)(b), obtained on heating and cooling with the corresponding applied pressures ranging from 0 GPa to 0.57 GPa. Thus, by subtracting the isobaric curves, the isothermal entropy change on loading (0$\rightarrow$p) and unloading (p$\rightarrow$0) can be obtained, as shown in Fig. \ref{fig:dS_signal}(c). In principle, the extra term $\Delta S_{vol}$ takes care of the differences observed in $\Delta S$ data in two different papers for the same plastic crystal (\textbf{NPG}) \cite{lloveras2019colossal,li2019colossal}. Nevertheless, it may be an indicative of an issue with the indirect method of obtaining $\Delta S$. In fact, the extra term $\Delta S_{vol}$ may not be necessary since calorimetric data should contain all the contributions to $\Delta S$.

\begin{figure}
    \centering
    \includegraphics[width=7.5cm,keepaspectratio]{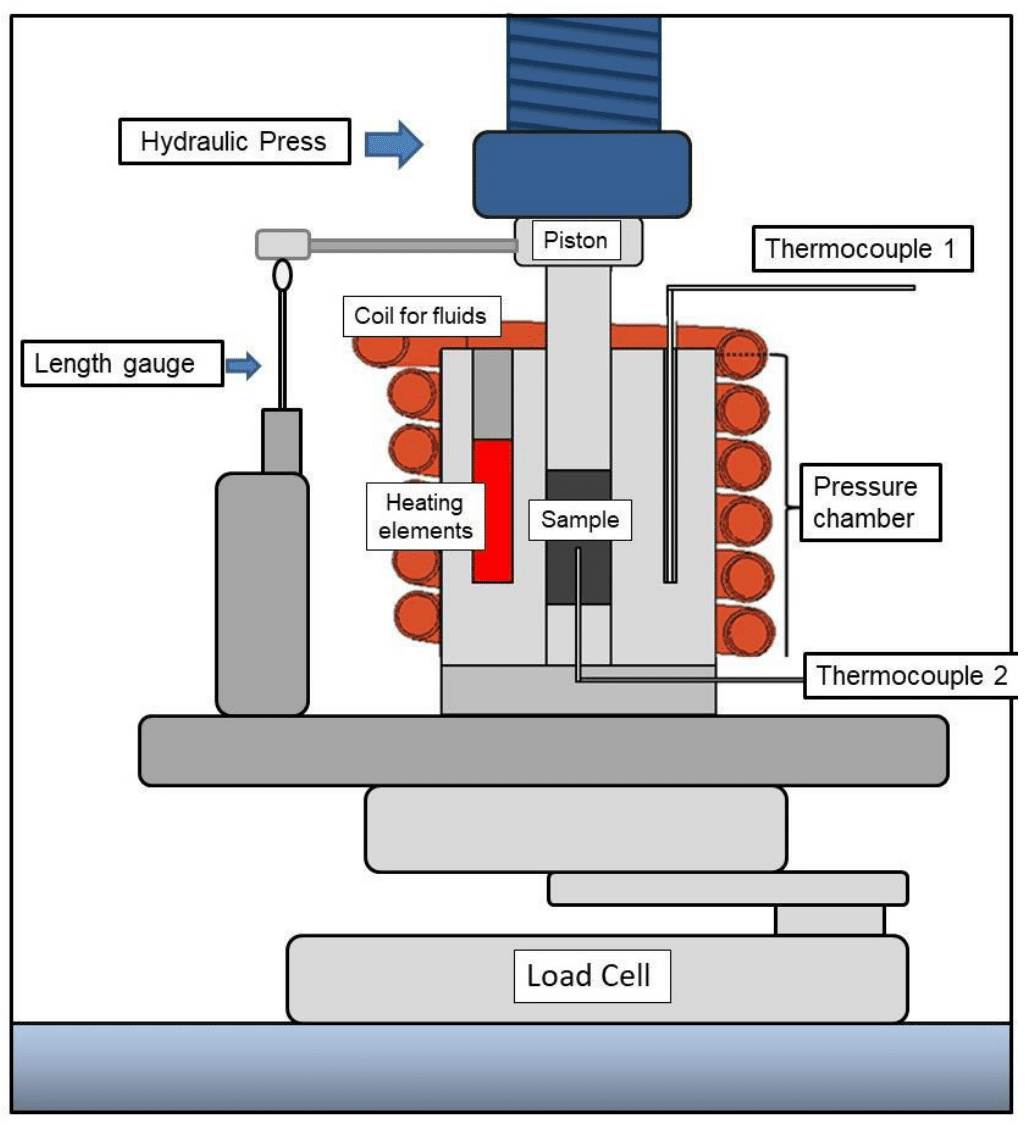}
    \caption{Experimental setup for the measurement of BC containing a cooling/heating system, pressure chamber, thermocouples to directly measure the temperature change on loading/unloading pressure, and a length gauge to acquire the strain curves $\gamma(T,p)$. Reprinted from reference \onlinecite{usuda2019giant} with permission of American Chemical Society.}
    \label{fig:measure1}
\end{figure}

\textbf{Direct measurements.}  A scheme of a homemade setup is depicted in Fig. \ref{fig:measure1}, where the pressure chamber consists of a cylindrical steel block, with a hole in the center to accommodate the sample and the moving piston. A thermocouple is coupled to the sample in order to monitor its temperature. The temperature of the system is changed using a thermostatic bath or liquid nitrogen. The load cell put underneath the apparatus registers the applied force; and the value of the pressure is obtained from standard equations. Note that the pressure is applied along one direction; however, according to equation \ref{poiseq}, for materials with the Poisson coefficient of 0.5, the uniaxial compression/decompression corresponds to an isotropic BC. In addition, from this experimental setup, the application/release of pressure occurs quasi-adiabatically and the adiabatic temperature change $\Delta T(p)$ of the sample is directly acquired.


\begin{figure}
    \centering
    \includegraphics[width=7cm,keepaspectratio]{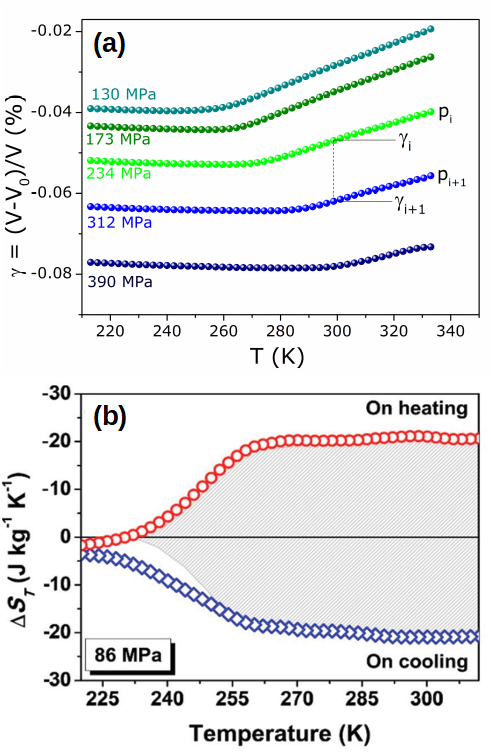}
    \caption{(a) Strain isobaric map $\gamma(T,p)$ measurements obtained by unloading and heating a nitrile butadiene rubber (\textbf{NBR}) elastomer using the homemade setup from reference \onlinecite{usuda2019giant} and (b) the calculated isothermal entropy change ($\Delta S$). The hatched area gives the reversible isothermal entropy change. Reprinted from reference \cite{usuda2019giant} with the permission of Americal Chemical Society. }
    \label{fig:strainmap}
\end{figure}

As can be seen in Fig. \ref{fig:measure1}, a length gauge is also connected to the piston to measure the sample strain during the compression/decompression and the heating/cooling processes. The data is acquired by increasing and reducing the sample temperature under constant stress to obtain a map of the strain $\gamma (T,P) = (V-V_0)/V_0$. An example of the obtained curves is given in Fig. \ref{fig:strainmap} for the Nitrile Butadiene Rubber (\textbf{NBR}) acquired by isobaric heating protocol \cite{usuda2019giant}. Thus, it is possible to indirectly obtain the entropy change through the Maxwell relation (see equation \ref{DeltaSS}). For this, the measurement should be taken by varying the temperature in a constant step $\Delta T = T_{i+1}-T_i$ at each pressure to numerically evaluate equation \ref{DeltaSS} as follows:

\begin{equation}\label{eqdisc}
\footnotesize
    \Delta S(T,\Delta p) = \frac{-V_0\Delta p}{2} \sum_{i=0}^{n}\left[\left(\frac{\Delta \gamma_i}{\Delta T_i}\right)_{p_i} + \left(\frac{\Delta \gamma_{i+1}}{\Delta T_i}\right)_{p_i+1}\right]
\end{equation}
where $\Delta p = p_{i+1}-p_i$ is the difference between two isobaric curves under consideration and $n$ is the number of points for each curve. For better visualization, all the quantities are represented in Fig. \ref{fig:strainmap}(a) and the resulting isothermal entropy change curves are given in Fig. \ref{fig:strainmap}(b) for the elastomer nitrile butadiene rubber (\textbf{NBR}). Nevertheless, it should be highlighted that $\Delta S(T,p)$ can only be obtained from the procedure using the method of Fig. \ref{fig:measure1} if the material presents a Poisson coefficient $\nu\sim 0.5$. For plastic crystals or other systems with $\nu \neq 0.5$, this method would result in the evaluation of the mechanocaloric, being neither elasto- or barocaloric effects.

Finally, Zhang et al. \cite{zhang2022colossal} published \textit{in-situ} measurements in carboranes using thermocouples and double-toroidal tungsten carbide anvil cells. The pressure/depressurization are performed through an automatic hydraulic pump to control different pressure increase rates.However, there is no complete information regarding the experimental methodology and, for this reason, it will not be approached here.

\section{Plastic barocaloric materials: recent findings}

\subsection{Plastic crystals}

\subsubsection{Motivation}

Recent findings for the colossal barocaloric effect in neopentylglycol (\textbf{NPG}) \cite{li2019colossal,lloveras2019colossal} have turned the community's attention to study plastic crystals. An adiabatic temperature change ($\Delta T_{ad}$) of 30 K for an applied pressure of 0.57 GPa was observed around the order-disorder transition for this material, as listed in Table \ref{materials}. This result is among the highest reported values for this class of material, accompanied by maximum isothermal entropy change ($\Delta S_{max}$) of 550 J/kg.K. This $\Delta S_{max}$ value is close to the one for R134a refrigerant of 520 J/kg.K \cite{mclinden2005thermophysical}, used nowadays in conventional gas-based refrigerators, as demonstrated by Lloveras et al. \cite{lloveras2019colossal}. A latter report on modified neopentane plastic crystal revealed even higher values of 690 J/kg.K and 600 J/kg.K for \textbf{AMP} and \textbf{TRIS}, respectively, for an applied pressure of 0.25 GPa~\cite{aznar2020reversible}; however, these are irreversible, which is a drawback for practical applications. More recently, Aznar et al. \cite{aznar2021reversible} observed a colossal \textbf{BC} response in 1-X-adamantane with X = Br and Cl (\textbf{Br-ada} and \textbf{Cl-ada}, respectively), where a reversible process occurs for these PCs around room temperature for a moderately applied pressure of 0.24 GPa with a working range temperature of $\sim$ 40 K, opening the possibility of using these materials in devices. However, each PC system presents its peculiarities and, for this reason, an overview of their barocaloric properties should be performed to organize the potential of these materials to be applied as refrigerant for solid-state cooling devices.  

\subsubsection{State of the art}

A list of the reported barocaloric properties in plastic crystals is given in Table \ref{materials}, where the maximum values were collected from the isothermal entropy change and adiabatic temperature change data. The listed values were acquired from the data obtained on heating and removing stress, except for \textbf{NPG$_1$}\cite{li2019colossal}, where the available data is only on cooling and loading. It is worth noting the discrepancies found for the caloric potentials of \textbf{NPG}\cite{lloveras2019colossal,li2019colossal}. There are some reasons for this occurrence: (i) Lloveras et al.~\cite{lloveras2019colossal} included corrections using X-ray diffraction data to analyze the entropy change; (ii) the use of approximation in equation \ref{deltfwff} can also be a reason and, finally, (iii) the use of different values for $C_p$ by those authors. For the barocaloric effect in plastic crystals, there is still a lack of a standard protocol to obtain the caloric potentials. 

A further study performed by the same group of Lloveras has shown that neopentane [C(CH$_3$)$_4$] derivatives also present great barocaloric potentials\cite{aznar2020reversible}. However, these have a melting temperature above room temperature for most of the materials given in Table \ref{materials}. Namely, the systems selected by Aznar et al. were: (NH$_2$)(CH$_3$)C(CH$_2$OH)$_3$ [2-amino-2-methyl-1,3-propanediol, \textbf{AMP}]; (NH$_2$)C(CH$_2$OH)$_3$ [2-amino-2(hydroxymethyl)propane-1,3-diol, \textbf{TRIS}]; (CH$_3$)C(CH$_2$OH)$_3$ [2-hydroxymethyl-2-methyl-1,3-propanediol, \textbf{PG}] and; (CH$_3$)$_3$C(CH$_2$OH) [2,2-dimethyl-1-propanol, \textbf{NPA}]. For the particular case of \textbf{NPA}, the working temperature for the barocaloric effect (T$_{FWHM}$) is higher than \textbf{NPG} for a comparable value of applied pressure. This might be due to the lowest transition temperature of \textbf{NPA}, where it becomes disorientated around room temperature and, consequently, the entropy curves become broader. Another set of PCs, presenting large \textbf{BC} responses are the adamantane (C$_{10}$H$_{16}$) with the substitution of one H atom by Cl (C$_{10}$H$_{15}$Cl, \textbf{Cl-ada}) and Br (C$_{10}$H$_{15}$Br, \textbf{Br-ada}) where a high T$_{FWHM}$ of $\sim$ 40 K is observed for an applied pressure of 0.24 GPa \cite{aznar2021reversible}. The slightly higher entropy change values for \textbf{Br-ada} is related to two successive ordered-disordered transitions from cubic to orthorhombic structure at 282 K, becoming a monoclinic phase at 308 K \cite{bazyleva2005thermodynamic}. The larger contribution rises from the room temperature transition, being the one depicted in Table \ref{materials}  \cite{aznar2021reversible} with the higher pressure-induced temperature transition rate ($dT_t/dP$) among the adamantanes.

\begin{table*}
\centering

\begin{threeparttable}

\begin{tabular}{c|c|c|c|c|c|c|c|c|c|c|c}
Material & Type & 
\begin{tabular}[c]{@{}c@{}}
$C_p$\\(J/kg K)
\end{tabular} & 
\begin{tabular}[c]{@{}c@{}}
$|\Delta T_{max}|$\\ (K)
\end{tabular} & 
\begin{tabular}[c]{@{}c@{}}
$|\Delta S_{max}|$\\ (J/kg K)
\end{tabular} & 
\begin{tabular}[c]{@{}c@{}}
$|\Delta P|$\\ (GPa)
\end{tabular} & 
\begin{tabular}[c]{@{}c@{}}
$T_{FWHM}$\\ (K)
\end{tabular} & 
\begin{tabular}[c]{@{}c@{}}
$T_t$\\ (K)
\end{tabular} & \begin{tabular}[c]{@{}c@{}}
$dT_t/dP$ \\(K/GPa)\end{tabular} &\begin{tabular}[c]{@{}c@{}}
$\Delta T_{hyst}$\\ (K)
\end{tabular} &\begin{tabular}[c]{@{}c@{}}
$P_{rev}$\\ (GPa)
\end{tabular} & Ref.\\

\hline\hline

\textbf{NPG$_1$}         & PC & 1940 & 64\tnote{a} & 390          & 0.09 & 15          &  313 & 130 & 18 &  & ~\onlinecite{li2019colossal}\\
\textbf{NPG$_2$}         &    &      & 10          & 410          & 0.09 & 17          &  313 & 130 & 24 & 0.15 & ~\onlinecite{lloveras2019colossal}\\
                &    &      & 42          & 550          & 0.57 & 48          &         &     & & & ~\onlinecite{lloveras2019colossal}\\
Neopentane \textbf{AMP}  & PC & 1790 & 15\tnote{b}          & 690\tnote{b}          & 0.24 & 23          &  353 & 80 & 50 &  & ~\onlinecite{aznar2020reversible}\\
Neopentane \textbf{TRIS} & PC & 1800 & 8\tnote{b}           & 600\tnote{b}          & 0.24 & 21          &  407 & 30 & 64 &  & ~\onlinecite{aznar2020reversible}\\
Neopentane \textbf{PG}   & PC & 1920 & 10          & 490          & 0.24 & 30          &  354 & 90 & 4 & 0.04 & ~\onlinecite{aznar2020reversible}\\
Neopentane \textbf{NPA}  & PC & 2790 & 42          & 470          & 0.59 & 61          &  232 & 170 & 35 & 0.19 & ~\onlinecite{aznar2020reversible}\\
\textbf{Cl-ada}          & PC &      & 48          & 220          & 0.24 & 44          &  245 & 270 & 10 & 0.03 & ~\onlinecite{aznar2021reversible}\\
\textbf{Br-ada}          & PC &      & 57          & 235          & 0.24 & 40          &  308 & 350 & 4 & 0.04 & ~\onlinecite{aznar2021reversible}\\
\textbf{\textit{p-}carb} &  PC  &  1590    & 19\tnote{a} & 98 & 0.06 & 20 &        308  & 380 & 11  & 0.01 & ~\onlinecite{zhang2022colossal}\\
\textbf{\textit{o-}carb} &  PC  &  1590 & 14\tnote{a} & 80 & 0.06   & 18 &         277 & 330 & 8  & 0.01 & ~\onlinecite{zhang2022colossal}\\
            &   &   & 9\tnote{d} &  & 0.30 &  &         &    &  & & ~\onlinecite{zhang2022colossal} \\
\textbf{\textit{m-}carb} &  PC  & 1590 & 13\tnote{a} & 72 & 0.06 & 19  &       286 & 330 & 9 & 0.01 &  ~\onlinecite{zhang2022colossal}\\

\hline \hline

\textbf{PDMS}            & EL & 1350 & 28\tnote{c}          & 140          & 0.39 &             & 150 &  & & & ~\onlinecite{carvalho2018giant}\\
\textbf{VNR}             & EL & 1930 & 25\tnote{c}          & 154          & 0.39 &             &  &     & & & ~\onlinecite{bom2018giant}\\
\textbf{ASR}             & EL &      & 41\tnote{c}          &             & 0.39 &             & 2 & 250 & & & ~\onlinecite{imamurasupergiant}\\
\textbf{NBR}             & EL & 1350 & 16\tnote{c}          & 68           & 0.39 &             &  &     & & & ~\onlinecite{usuda2019giant}\\
\textbf{PU}             & EL & 1350 & 13\tnote{c}          & 96           & 0.22 &             &  &     & & & ~\onlinecite{bocca2021giant}\\
\hline\hline
\end{tabular}

\begin{tablenotes}
\small
\item[a] calculated using $|\Delta T| = T |\Delta S|/C_P$.
\item[b] Irreversible BC ~\cite{aznar2020reversible}.
\item[c] measured directly with the setup of Figure \ref{fig:measure1}.
\end{tablenotes}

\end{threeparttable}

\caption{Barocaloric response and physical parameters of plastic crystals (PC) and elastomers (EL). $C_p$: heat capacity at room temperature, $|\Delta T_{max}|$: maximum adiabatic temperature change, $|\Delta S_{max}|$: maximum isothermal entropy change, $|\Delta P|$: applied pressure, $T_{FWHM}$: full width at half maximum of the entropy curve, $T_t$: transition temperature, being the melting for PC and glass transition temperature for EL at ambient pressure, $dT_t/dP$: shift of the temperature change under applied pressure obtained on cooling, $\Delta T_{hyst}$: thermal hysteresis obtained from DSC measurements and P$_{rev}$ which correspond to the minimum applied pressure to achieve reversible BC. The values were collected for $|\Delta S_{max}|$ and $|\Delta T_{max}|$ curves obtained by unloading (p$\to$atmospheric pressure) and heating.}
\label{materials}
\end{table*}

Considering practical use of barocaloric materials, greater $dT_t/dP$ values are desirable since it is related to higher working range temperature for the BC with lower hysteresis losses. More recently, Zhang et al. \cite{zhang2022colossal} obtained high $dT_t/dP$ values for three positional isomers of carborane (C$_2$B$_{10}$H$_{12}$), namely: ortho-carborane \textbf{(o-carb)}, meta-carborane \textbf{(m-carb)} and para-carborane \textbf{(p-carb)}. The three materials present an orthorhombic to tetragonal structural transition and the distinction between these carboranes relies on the B occupation in the structure, where the highest symmetry occurs when B are positioned at the icosahedron opposite vertices (p-carb). As given in Table \ref{materials}, this group of PC also presents low thermal hysteresis ($T_{hyst}$), with p-carb being the one with improved BC responses due to its higher symmetry. In addition, the minimum applied pressure required for reversibility ($P_{rev}$), where there is an overlap between the entropy change curves on heating and on cooling, is the lowest for carboranes among the PCs reported in the literature. The improved BC response of p-carb is related to the higher symmetry over o-carb and m-carb where boron occupies the vertices of the icosahedron formed by the atoms. Although reducing hysteresis losses are desirable for practical applications, the carboranes present low entropy change values. Moreover, the temperature change obtained from \textit{in-situ} measurements is also obtained by the authors and, as can be noted, is below the calculated from the entropy change (see Table \ref{materials}). This is due to the lack of thermal insulation during the measurement; however, there is no complete information on the procedure in Ref. \cite{zhang2022colossal}.  In this regard, Br-ada BC properties are favorable for its use as a refrigerant, considering it presents the lowest $T_{hyst}$. Thus, further investigation on how to improve symmetry and successive phase transitions might be a path to select the best PC.

As can be noted from Table \ref{materials}, the broad range of applied pressures used by different authors makes difficult the comparison of the BC among materials. In order to deepen this analysis, we normalized both entropy change and temperature change by the applied pressure: $|\Delta S_{max}/\Delta P|$ and $|\Delta T_{max}/\Delta P|$. For a reasonable comparison between the properties of these materials, the data were collected for applied pressures around 0.1 GPa, taking into account the values related to reversible processes and might be underestimated. Except for \textbf{AMP} and \textbf{TRIS}, for which the processes are irreversible for this range of applied pressure \textbf{as highlighted in Table \ref{materials}}. Materials presenting large $|\Delta S_{max}/\Delta P|$, with the highest associated temperature change $|\Delta T_{max}/\Delta P|$, are better for barocaloric applications. It becomes clear that plastic crystals present the largest normalized entropy changes in comparison to the elastomers. In particular, \textbf{NPA} and \textbf{NPG$_2$} are the ones with better properties. However, as discussed below, for practical applications, the process should be reversible and, for \textbf{NPA}, applied pressure above 0.30 GPa is required for the reversibility. In this case, among the plastic crystals, we should classify \textbf{NPG$_2$} as the material with the best reversible BC around this range of applied pressure.

\begin{figure}
    \includegraphics[width=8.5cm,keepaspectratio]{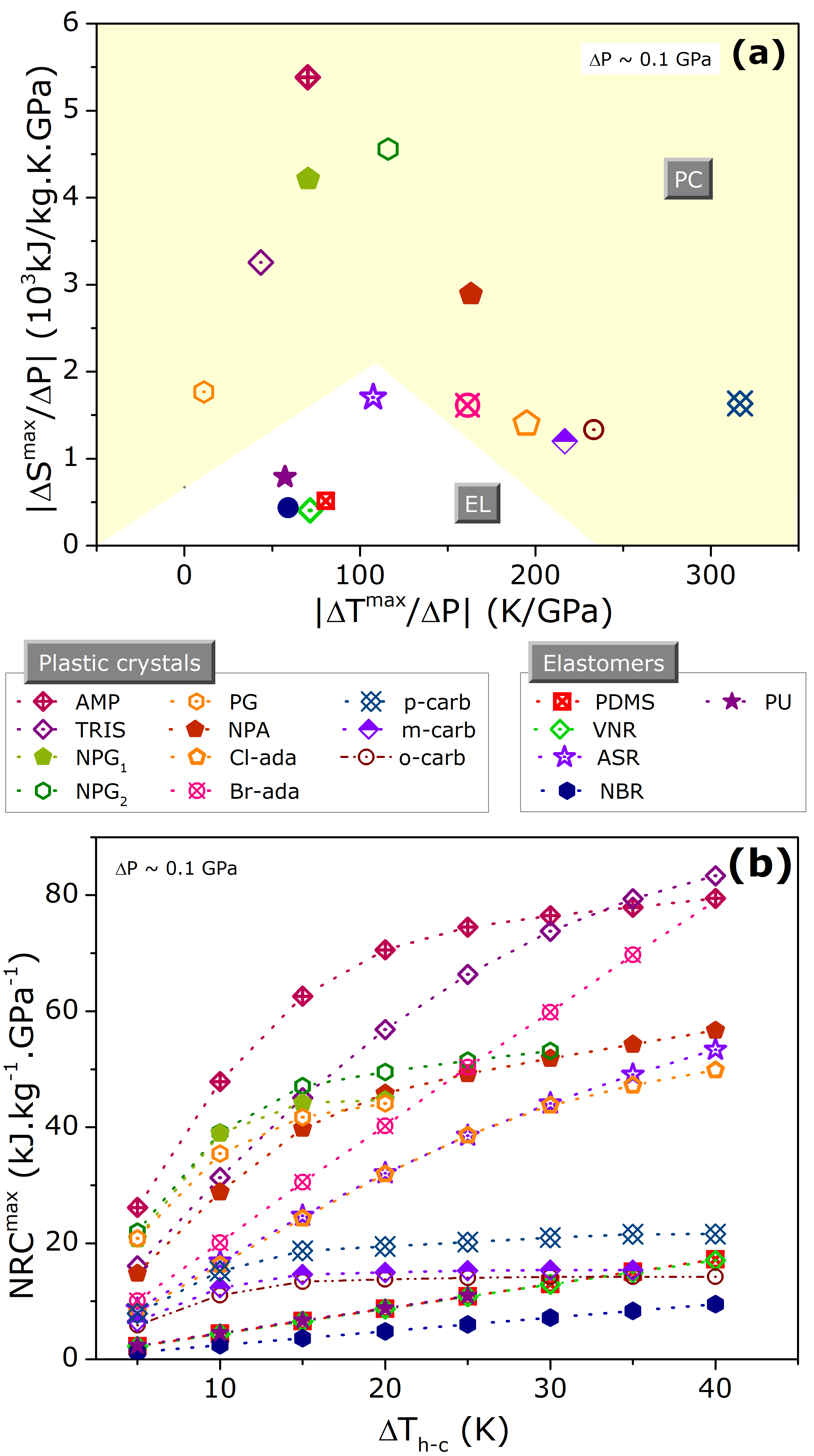}
    \caption{\textbf{(a)} Normalized $|\Delta S^{max}/\Delta P|$ \textit{vs} $|\Delta T/\Delta P|$ values for plastic crystals (yellow region) and elastomers (green region). This diagram was made for an applied pressure of ca. 0.1 GPa and the materials highlighted with $^*$ corresponding to irreversible processes. \textbf{(b)} Normalized maximum Refrigerant Capacity (NRC$^{max}$) \textit{vs} $\Delta T_{h-c} = T_{hot} - T_{cold}$ (hot and cold reservoir temperature difference) for plastics crystals and elastomers. All the values presented here were acquired from the curves obtained on heating and unloading. The values of applied pressure are: \textbf{\textbf{o-,m-,p-carb} $|\Delta P|$ = 0.06 GPa;} \textbf{NPA} $|\Delta P|$ = 0.08 GPa; \textbf{ASR}, \textbf{PDMS}, \textbf{VNR}, \textbf{NBR}\textbf{, \textbf{PU},} and \textbf{NPG$^{1,2}$} $|\Delta P|$ = 0.09 GPa; \textbf{AMP}, \textbf{PG}, \textbf{Br-ada} and \textbf{Cl-ada} $|\Delta P|$ = 0.10 GPa; \textbf{TRIS} $|\Delta P|$ = 0.11 GPa.}
    \label{fig:baro_mat}
\end{figure}

For a further analysis of these materials, we have calculated the normalized maximum refrigerant capacity:
\begin{equation}
NRC^{max} = \frac{|\int_{T_{hot}}^{T_{cold}}\Delta SdT|}{\Delta P}\label{xyz}
\end{equation}
obtained from the entropy curves around 0.1 GPa, as a function of the temperature difference between cold and hot reservoirs. Similar to the previous graph, the entropy curves for the NRC calculation were acquired from heating and releasing pressure curves, except for \textbf{NPG$_1$} data. Given the reduced temperature range of reversible processes, the NRC was acquired from the irreversible entropy change curves to extend the $\Delta T_{h-c}$ to 60 K for most systems, as shown in figure \ref{fig:baro_mat}(b). Although the plastic crystals present the higher NRC$^{max}$ values among all the presented materials, as $\Delta T_{h-c}$ increases, the curves tend to saturate. As an example, see the corresponding curves for \textbf{NPG$_2$} and \textbf{NPA}, which from Fig. \ref{fig:baro_mat}(a) presents the higher normalized entropy and temperature changes. The saturation of NRC$^{max}$ reveals the limitation on the barocaloric performance of these materials. In contrast, although presenting lower normalized entropy change, the NRC$^{max}$ of \textbf{Br-ada} overcomes all the PCs at $\Delta T_{h-c}$ = 40 K, revealing that it can be used in devices that require a large range of temperature operations. Nevertheless, distinctly from the observations of figure \ref{fig:baro_mat}(a), the materials presenting the highest NRC$^{max}$ values are \textbf{AMP}, \textbf{TRIS} and \textbf{Br-ada}. Although these seem to be the best materials, we should emphasize that \textbf{AMP} and \textbf{TRIS} do not present reversibility across the transition, which is an issue for practical applications.

\subsubsection{Microscopic scenario}

The isothermal entropy change $\Delta S$ for \textit{plastic crystals} has two main contributions during transition: \textit{(i)} volumetric ($\Delta S_V$) and \textit{(ii)} configurational ($\Delta S_C$) terms ~\cite{lloveras2019colossal,aznar2020reversible}. Through X-ray diffraction and calorimetry data\cite{aznar2020reversible}, Aznar et al. calculated both contributions for neopentane \textbf{AMP}, \textbf{TRIS}, \textbf{PG} and \textbf{NPA} and adamantine \textbf{Br-ada} and \textbf{Cl-ada} plastic crystals to compare with the endothermic entropy change obtained during the materials disordered-ordered transition at atmospheric pressure, denoted as $\Delta S_{d\rightarrow o}(p_o)$ in Table \ref{tab:contributons}. The volumetric contribution (due to lattice distortions) was calculated considering \cite{aznar2020reversible}:
\begin{equation}\label{eq:dSv}
\Delta S_V \approx \frac{\langle\alpha\rangle}{\langle\kappa_V\rangle}\Delta V_t    
\end{equation}
\noindent where $\langle\alpha\rangle
$ and $\langle\kappa_V\rangle$ are the average thermal expansion and average isothermal compressibility, respectively, for the two phases, and $\Delta V_t$ is the volumetric change across the phase transition. Equation \ref{eq:dSv} is an approximation and must be used with caution \cite{jenau1996crystal}. This evaluation found that the fraction of $\Delta S_V$ is around 15\% for \textbf{TRIS}, \textbf{PG} and \textbf{AMP} neopentane and 40\% for \textbf{NPA} in relation to the total entropy obtained from DSC measurements \cite{aznar2020reversible}. For the particular case of \textbf{Br-ada}, the high volumetric entropy change contribution of 73\% rises from the remaining orthorhombic phase across the disordered monoclinic to ordered cubic transition. This finding reveals that the main contribution for the large BC observed in these materials comes from the configurational entropy. It is worth mentioning that the volumetric contribution for the \textbf{NPA} material is due to the high unit cell volume of the triclinic phase at low temperatures~\cite{salud1999two}. For the configurational entropy, Raman spectroscopy analysis was used and this term was estimated through the relation:
\begin{equation} \label{eq:dSc}
    \Delta S_C = \frac{R}{M}\ln\left(\frac{N_I}{N_{II}}\right)
\end{equation} 
\noindent with $R$ standing for the universal gas constant, $M$ the molar mass and $N_I$ and $N_{II}$ the number of molecular conformations in the disordered and ordered phases, respectively. All the calculated values obtained by Aznar et al. are listed on Table \ref{tab:contributons}, along with their corresponding fractions related to the total value of $\Delta S_{exo}(p_0)$ obtained at atmospheric pressure ($p_0$). In particular, for \textbf{NPA}, the fraction of $\Delta S_C$ is more than 50\% larger than the experimental value of the entropy. This is because the orientational disorder for these molecular arrangements is smaller than predicted. Although the calculated values show that the main contribution to the entropy change arises from the molecular contribution, these results are overestimated due to the inaccessibility to all possible configurations. Therefore, a more accurate understanding of the microscopic origins of the entropy still lacks in the literature.
\begin{table}
\resizebox{\columnwidth}{!}{%
\begin{tabular}{c|c|cc|cc}
 Material & \begin{tabular}[c]{@{}c@{}} $\Delta S_{d\rightarrow o}(p_o)$\\ (J/kg.K)\end{tabular} & \begin{tabular}[c]{@{}c@{}}\small$\Delta S_V$\\ (J/kg.K)\end{tabular} & \begin{tabular}[c]{@{}c@{}} Frac.\\ (\%)\end{tabular} & \begin{tabular}[c]{@{}c@{}}$\Delta S_C$\\ (J/kg.K)\end{tabular} & \begin{tabular}[c]{@{}c@{}} Frac.\\ (\%)\end{tabular} \\ \hline \hline
\textbf{PG}       & 485    & 80    & 13     & 325    & 67     \\
\textbf{NPA}      & 204    & 52    & 39     & 321    & 157    \\
\textbf{TRIS}     & 682    & 90    & 13     & 287    & 42     \\
\textbf{AMP}      & 632    & 100   & 16     & 315    & 50    \\ 
\textbf{Cl-ada}      & 136    & 51   & 37    & 112    & 82    \\ 
\textbf{Br-ada}      & 104    & 76   & 73     & 62    & 60 \\
\hline \hline
\end{tabular}%
}
\caption{Entropy change across the disordered to ordered phase transition at atmospheric pressure [$\Delta S_{d \rightarrow o } (p_o)$]; and the volumetric ($\Delta S_V$) and configuration ($\Delta S_C$) entropy changes calculated through equations \ref{eq:dSv} and \ref{eq:dSc}, respectively. This table also presents the corresponding fractions of each entropy change term.} \label{tab:contributons}
\end{table}


\subsubsection{Drawbacks}

Aznar et al.~\cite{aznar2020reversible} discussed the reversibility of the BC on the neopentane plastic crystals.
Two modified neopentane plastic crystals with the substitution of the methyl groups were chosen for the study: amino (NH$_2$) and hydroxymethyl (CH$_2$OH). The barocaloric evaluation revealed that the reversible response is around 10-20 K lower than that for \textbf{PG} and \textbf{NPA} plastic crystals, belonging to the hydroxymethyl group. The reversible entropy change ($\Delta S_{rev}$) condition is given by overlapping the isothermal entropy change curves, independently obtained on heating and on cooling.

In other words, the exothermic and endothermic transition temperatures should be equal under an applied pressure - this is denoted as reversal pressure. Consequently, due to hysteresis losses of the first-order phase transition, $\Delta S_{irr} > \Delta S_{rev}$, as can be observed from Table \ref{materials} for \textbf{NPA} and \textbf{PG} materials. In contrast, \textbf{TRIS} and \textbf{AMP} present a large hysteresis at low pressures, turning the minimum pressure for the reversibility analysis unfeasible~\cite{aznar2020reversible}. Thus, for the mentioned plastic crystals, the BC is irreversible. As can be seen from Table \ref{materials}, the working temperature for the barocaloric effect is also reduced for the reversible effect, being around 20 K lower for NPA plastic crystal. 
\subsection{Elastomers}

\subsubsection{Motivation}

Elastomers is another family of materials presenting an extraordinary large barocaloric response. Although a temperature change of $\sim$ 12 K was observed in 1942 by stretching natural rubbers (NR)~\cite{dart1942rise}, only in recent years the mechanocaloric effect of these materials have been explored systematically for solid-state cooling application~\cite{miliante2020unveiling,bom2017note}. The evaluations reported in the literature revealed great barocaloric potentials for the elastomers, which are interesting given the absence of structural transitions for most of these materials. Such behavior has its origin at the molecular level; the rearrangement of the polymeric chains during releasing/applying pressure is the main mechanism for the barocaloric effect, as recently demonstrated by Miliante et al.~\cite{miliante2020unveiling}. 

\subsubsection{State of the art}

Among the results reported in the literature, elastomers like acetoxy silicone rubber (\textbf{ASR})~\cite{imamurasupergiant} and polydimethylsiloxane (\textbf{PDMS})~\cite{carvalho2018giant} present adiabatic temperature changes comparable with the observed in plastic crystals, as can be seen in Table \ref{materials}. As can be noted, the barocaloric response of \textbf{ASR} occurs in the vicinity of the PCs response, as can be seen in Figs. \ref{fig:baro_mat}(a) and (b); revealing its great potential for solid-state cooling technologies. The supergiant BC of \textbf{ASR} is due to a crystalline-amorphous transition that occurs around 250 K and, combined with the molecular rearrangement, leads to a measured $\Delta T$ of ~40 K for an applied pressure of 0.39 GPa. Above the glass transition temperature, a $\Delta T$ of 30 K is observed, which is only due to the amorphous phase contribution. For this reason, that compound is the one presenting the best BC potentials compared with the other amorphous materials, such as \textbf{VNR} and \textbf{NBR}. For the \textbf{PDMS}, as can be seen from Table \ref{materials}, the T$_G$ is far from room temperature, and $\Delta T$ of 28 K occurs for $\Delta P$ of 0.39 GPa. Although this value is below the observed for \textbf{ASR}, it is comparable with those estimated for plastic crystals, such as neopentane \textbf{PG} and \textbf{AMP}. More importantly, the BC for \textbf{ASR} is reversible at low applied pressures ($\Delta P < 0.2$ GPa), which is an important feature concerning practical applications. Although these results seems promising, few reports have been dedicated to these investigations. 

Most of the barocaloric potentials for elastomers do not present a well-defined peak. This is due to the high melting point of rubbers (well above 400 K) with the exception of \textbf{PDMS}. Given this, the $\Delta S$ and $\Delta T$ curves tend to present high values in a large temperature range. For instance, the reversible entropy change of \textbf{ASR} has values above 150 J/kg.K in a temperature range of $\sim$ 30 K for an applied pressure of 0.17 MPa. While for \textbf{NBR}, a table-like behavior occurs between 255 K and 315 K for 0.086 MPa of applied pressure. The temperature range of 40 K for an entropy change of -40 J/kg.K for \textbf{ASR} can be overestimated since the limit for the experimental setup is 320 K. In this sense, it was not possible to obtain the temperature span (T$_{FWHM}$) for this class of materials, being the responsible for the absence of these values on Table \ref{materials}. For the reasons stated above the values depicted in Fig. \ref{fig:baro_mat}(a) are related to the global maximum of the curves available in literature normalized for an applied pressure of $\sim$ 0.1 GPa. Although the elastomers are in the lower region of the chart, the maximum adiabatic temperature change for \textbf{NBR}, \textbf{VNR} and \textbf{PDMS} and polyurethane (\textbf{PU}) are in the same range as \textbf{TRIS} and \textbf{AMP}. While \textbf{ASR}, as already mentioned, presents a larger $|\Delta T^{max}/\Delta P|$ that is $\sim$ 10 K/GPa below \textbf{NPG$_2$}, revealing its great \textbf{BC} response among the evaluated elastomers. 

A better approach to compare the barocaloric potential between plastic crystals and elastomers is through the NRC$^{max}$ \textit{versus} $\Delta T_{h-c}$ curves, shown in Figure \ref{fig:baro_mat}(b). As previously mentioned, the plastic crystals tend to a saturation above $\Delta T_{h-c}\approx20$ K, whereas a linear increase of NRC$_{max}$ for the elastomers occurs, as seen for \textbf{ASR}, \textbf{PDMS}, \textbf{VNR} and \textbf{NBR} (for this range of $\Delta T_{h-c}$). This behavior is an advantage for applications of the elastomers, since it can be used in a wide range of temperatures and at relatively low applied pressures (0.1 GPa). In addition, the reversibility of the BC for the elastomers is appealing for solid-state cooling technology.

\subsubsection{Microscopic scenario}

From a fundamental point of view, the mechanism responsible for such a great barocaloric effect is still under study. The best explanation so far is that these outstanding results arise from a reduction of the free volume and number of accessible vibrational/rotational states of the molecules due to the applied pressure. In contrast to the plastic crystals, where only rotations of the molecules are possible, polymers present intra/intermolecular interactions with their rotational and vibrational modes available. For the elastomers, the configurational contribution is associated to the volumetric change, since the molecular interactions are pressure and volume dependent. In simple words, the main contributions for the isothermal entropy change $\Delta S$ for this class of materials are: \textit{(i)} the potential energy contribution, comprising the inter- and intra-molecular interactions; and, \textit{ii)} volumetric change during compression/decompression, $(\partial V/\partial P)_T$\cite{miliante2020unveiling}. The intramolecular contribution concerns the bond stretching and angle deformation during compression, with the addition of a torsion from the C-C/C-H bonds in the polymer, a dihedral energy; while the intermolecular interactions ($\Delta S_{inter}$) comprises the Coulomb and van der Walls forces between a pair of molecules. Note, that both terms in the potential energy are volume dependent.

Through molecular dynamics simulations, combined with thermodynamic analysis, Miliante et al.~\cite{miliante2020unveiling} have elucidated that the origin of BC in non-vulcanized natural rubbers (NR) relies on an unusual reduction of the intermolecular potential energy~\cite{miliante2020unveiling}. Under an adiabatic process, the work performed on the system by the compressing force must be equal to the variation of the internal energy of the system. This last comprises kinetic energy (related to the temperature) and potential energy (the sum of intramolecular and intermolecular energies). Potential energy is governed by atomic interactions, therefore atomic motion rules how much of the external energy, supplied as work, can be converted into kinetic energy. From simulated atomic trajectories, the contribution of each energy component was computed for the natural rubber/cis-1,4-polyisopropene during compression and decompression cycles: $U(T,V;P) = U_{pot} + U_{kin}$. Thus, using the simulation results under isothermal conditions, the entropy change could be calculated through the first law of thermodynamic $dU = dW = TdS - PdV$, as follows:

\begin{equation}
    \Delta S_{T} = \frac{1}{T}\left[\int \left( \frac{\partial U}{\partial P}\right)_T dP + \int P \left(\frac{\partial V}{\partial P}\right)_T dP \right]. \label{eq:dScalc}
\end{equation}

\begin{figure}
    \centering
   \includegraphics[width=7.5cm,keepaspectratio]{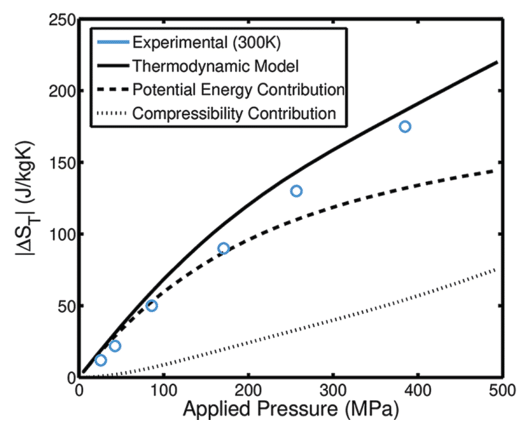}
    \caption{Simulated and experimental isothermal entropy change as a function of the applied pressure for the natural rubber \textit{cis}-1,4 polysoprene. The potential energy and compressibility contribution, from equation \ref{eq:dScalc}, are given separately where it is possible to observe that the inter/intramolecular interactions are the main responsible for the BC in NR. Reprinted from reference \cite{miliante2020unveiling} with permission of American Chemical Society.}
    \label{exp_teo}
\end{figure}

The result of these simulations are given in figure \ref{exp_teo}, showing a good agreement with the experimental pressure dependence of the entropy change for NR at 300 K. As can be noted, the potential contribution is the main term contributing to the large BC observed in the elastomer. The main responsible for this observation is the reduction of the energy related to the intermolecular interactions, namely, van der Walls and Coulomb energies. As pressure increases, intermolecular energy decreases, which enhances the gain in kinetic energy perceived macroscopicaly, and temperature increases. When submitted to higher pressures, the molecules approach each other, and at a given point the intermolecular interaction become repulsive. Under these conditions, the intermolecular energy increases, reducing the possible gain in kinetic energy.

\subsubsection{Drawbacks}

Some of these elastomers crystallize under high stress values. For uniaxial elongation, applied stress of the order of 200-400\% of strain crystallizes the NR, as shown by Toki et al. ~\cite{toki2000strain}. In the case of acetoxy silicone rubber (\textbf{ASR}), a partial amorphous to crystalline transition at reduced temperatures (ca. 200 K at atmospheric pressure), is responsible for a colossal BC response~\cite{imamurasupergiant}. For these cases, the rubber tends to maintain its configuration and may not return to its initial condition in an irreversible process. In another words, under this condition, the plastic behavior of the rubbers are drastically reduced being one of the reasons for the appearance of mechanical hysteresis~\cite{le2017energy}. For this reason, a throughout evaluation of materials performances under a large range of mechanical stimuli is required to identify the threshold for applied pressure, focusing on applications of these materials~\cite{aznar2020reversible}.

\section{Concluding Remarks}

The colossal barocaloric effect in neopentylglycol (\textbf{NPG}), with a maximum adiabatic temperature change of the order of 30 K and a maximum isothermal entropy change of 550 J/kgK, for 0.57 GPa of applied pressure change, opened a large avenue of research in caloric effects for plastic crystals. It is worth saying that  these astonishing values are very close to the isothermal entropy change for R134a refrigerant, used nowadays in conventional gas-based refrigerators. In addition to plastic crystals, we can also add to this amazing family the elastomers (see Materials Survey section). 

For a fair comparison of these materials, we have normalized both entropy change and temperature change by the applied pressure: $|\Delta S_{max}/\Delta P|$ and $|\Delta T_{max}/\Delta P|$. Materials with high values of $|\Delta S_{max}/\Delta P|$, and, simultaneously, high values of $|\Delta T_{max}/\Delta P|$, are better for barocaloric applications. Plastic crystals present the largest normalized entropy changes in comparison to the elastomers, being \textbf{NPA} and \textbf{NPG$_2$} the ones with better properties. For a further analysis, we have also calculated the normalized maximum refrigerant capacity NRC$^{max}$ - defined as in equation \ref{xyz} and obtained from the entropy curves around 0.1 GPa. Although plastic crystals present the higher NRC$^{max}$ values among all the presented materials, as the integration temperature width $\Delta T_{h-c}$ increases, those curves tend to saturate. This fact reveals the limitation on the barocaloric performance for plastic crystals. In contrast, although presenting lower normalized entropy change, the NRC$^{max}$ for the elastomer \textbf{Br-ada} overcomes all the PCs at $\Delta T_{h-c}$ = 40 K, revealing that it can be used in devices that require a large range of temperature operations. 

Thus, in this review, we have presented a brief survey about plastic crystals and elastomers, as well as the thermodynamic for mechanocaloric effects. The most important experimental techniques were also described, highlighting possible divergences of analysis, such as the addition of an extra entropy contribution obtained by X-ray diffraction. We also delivered a comprehensive literature overview, for both families of compounds, plastic crystals and elastomers. Along these discussions, we have presented the state of the art, microscopic analysis for the entropy change and possible drawbacks for the field. \\



\section*{Acknowledgements}
VMA thanks FCT/MCTES and IFIMUP for the portuguese funds in the framework of the projects UIDB/50006/2020 and UIDP/50006/2020. MSR thanks the CNPq, FAPERJ, INCT of \textit{Refrigera\c{c}\~{a}o e Termof\'{i}sica}. VF acknowledges the support of US/JUNTA/FEDER-UE (grant US-1260179), Consejería de Economía, Conocimiento, Empresas y Universidad de la Junta de Andalucía (grant P18-RT-746) and and AEI/FEDER-UE (grant PID2019-105720RB-I00)
.\\

\bibliography{bib.bib}

\begin{thebibliography}{34}%
\makeatletter
\providecommand \@ifxundefined [1]{%
 \@ifx{#1\undefined}
}%
\providecommand \@ifnum [1]{%
 \ifnum #1\expandafter \@firstoftwo
 \else \expandafter \@secondoftwo
 \fi
}%
\providecommand \@ifx [1]{%
 \ifx #1\expandafter \@firstoftwo
 \else \expandafter \@secondoftwo
 \fi
}%
\providecommand \natexlab [1]{#1}%
\providecommand \enquote  [1]{``#1''}%
\providecommand \bibnamefont  [1]{#1}%
\providecommand \bibfnamefont [1]{#1}%
\providecommand \citenamefont [1]{#1}%
\providecommand \href@noop [0]{\@secondoftwo}%
\providecommand \href [0]{\begingroup \@sanitize@url \@href}%
\providecommand \@href[1]{\@@startlink{#1}\@@href}%
\providecommand \@@href[1]{\endgroup#1\@@endlink}%
\providecommand \@sanitize@url [0]{\catcode `\\12\catcode `\$12\catcode
  `\&12\catcode `\#12\catcode `\^12\catcode `\_12\catcode `\%12\relax}%
\providecommand \@@startlink[1]{}%
\providecommand \@@endlink[0]{}%
\providecommand \url  [0]{\begingroup\@sanitize@url \@url }%
\providecommand \@url [1]{\endgroup\@href {#1}{\urlprefix }}%
\providecommand \urlprefix  [0]{URL }%
\providecommand \Eprint [0]{\href }%
\providecommand \doibase [0]{https://doi.org/}%
\providecommand \selectlanguage [0]{\@gobble}%
\providecommand \bibinfo  [0]{\@secondoftwo}%
\providecommand \bibfield  [0]{\@secondoftwo}%
\providecommand \translation [1]{[#1]}%
\providecommand \BibitemOpen [0]{}%
\providecommand \bibitemStop [0]{}%
\providecommand \bibitemNoStop [0]{.\EOS\space}%
\providecommand \EOS [0]{\spacefactor3000\relax}%
\providecommand \BibitemShut  [1]{\csname bibitem#1\endcsname}%
\let\auto@bib@innerbib\@empty
\bibitem [{\citenamefont {Ismail}\ \emph {et~al.}(2021)\citenamefont {Ismail},
  \citenamefont {Yebiyo},\ and\ \citenamefont {Chaer}}]{ismail2021review}%
  \BibitemOpen
  \bibfield  {author} {\bibinfo {author} {\bibfnamefont {M.}~\bibnamefont
  {Ismail}}, \bibinfo {author} {\bibfnamefont {M.}~\bibnamefont {Yebiyo}},\
  and\ \bibinfo {author} {\bibfnamefont {I.}~\bibnamefont {Chaer}},\ }\bibfield
   {title} {\bibinfo {title} {A review of recent advances in emerging
  alternative heating and cooling technologies},\ }\href@noop {} {\bibfield
  {journal} {\bibinfo  {journal} {Energies}\ }\textbf {\bibinfo {volume}
  {14}},\ \bibinfo {pages} {502} (\bibinfo {year} {2021})}\BibitemShut
  {NoStop}%
\bibitem [{\citenamefont {Miliante}\ \emph {et~al.}(2020)\citenamefont
  {Miliante}, \citenamefont {Christmann}, \citenamefont {Usuda}, \citenamefont
  {Imamura}, \citenamefont {Paix{\~a}o}, \citenamefont {Carvalho},\ and\
  \citenamefont {Muniz}}]{miliante2020unveiling}%
  \BibitemOpen
  \bibfield  {author} {\bibinfo {author} {\bibfnamefont {C.~M.}\ \bibnamefont
  {Miliante}}, \bibinfo {author} {\bibfnamefont {A.~M.}\ \bibnamefont
  {Christmann}}, \bibinfo {author} {\bibfnamefont {E.~O.}\ \bibnamefont
  {Usuda}}, \bibinfo {author} {\bibfnamefont {W.}~\bibnamefont {Imamura}},
  \bibinfo {author} {\bibfnamefont {L.~S.}\ \bibnamefont {Paix{\~a}o}},
  \bibinfo {author} {\bibfnamefont {A.~M.~G.}\ \bibnamefont {Carvalho}},\ and\
  \bibinfo {author} {\bibfnamefont {A.~R.}\ \bibnamefont {Muniz}},\ }\bibfield
  {title} {\bibinfo {title} {Unveiling the origin of the giant barocaloric
  effect in natural rubber},\ }\href@noop {} {\bibfield  {journal} {\bibinfo
  {journal} {Macromolecules}\ }\textbf {\bibinfo {volume} {53}},\ \bibinfo
  {pages} {2606} (\bibinfo {year} {2020})}\BibitemShut {NoStop}%
\bibitem [{\citenamefont {Das}\ \emph {et~al.}(2020)\citenamefont {Das},
  \citenamefont {Mondal},\ and\ \citenamefont {Reddy}}]{das2020harnessing}%
  \BibitemOpen
  \bibfield  {author} {\bibinfo {author} {\bibfnamefont {S.}~\bibnamefont
  {Das}}, \bibinfo {author} {\bibfnamefont {A.}~\bibnamefont {Mondal}},\ and\
  \bibinfo {author} {\bibfnamefont {C.~M.}\ \bibnamefont {Reddy}},\ }\bibfield
  {title} {\bibinfo {title} {Harnessing molecular rotations in plastic
  crystals: a holistic view for crystal engineering of adaptive soft
  materials},\ }\href@noop {} {\bibfield  {journal} {\bibinfo  {journal}
  {Chemical Society Reviews}\ }\textbf {\bibinfo {volume} {49}},\ \bibinfo
  {pages} {8878} (\bibinfo {year} {2020})}\BibitemShut {NoStop}%
\bibitem [{\citenamefont {Lloveras}\ \emph {et~al.}(2019)\citenamefont
  {Lloveras}, \citenamefont {Aznar}, \citenamefont {Barrio}, \citenamefont
  {Negrier}, \citenamefont {Popescu}, \citenamefont {Planes}, \citenamefont
  {Ma{\~n}osa}, \citenamefont {Stern-Taulats}, \citenamefont {Avramenko},
  \citenamefont {Mathur} \emph {et~al.}}]{lloveras2019colossal}%
  \BibitemOpen
  \bibfield  {author} {\bibinfo {author} {\bibfnamefont {P.}~\bibnamefont
  {Lloveras}}, \bibinfo {author} {\bibfnamefont {A.}~\bibnamefont {Aznar}},
  \bibinfo {author} {\bibfnamefont {M.}~\bibnamefont {Barrio}}, \bibinfo
  {author} {\bibfnamefont {P.}~\bibnamefont {Negrier}}, \bibinfo {author}
  {\bibfnamefont {C.}~\bibnamefont {Popescu}}, \bibinfo {author} {\bibfnamefont
  {A.}~\bibnamefont {Planes}}, \bibinfo {author} {\bibfnamefont
  {L.}~\bibnamefont {Ma{\~n}osa}}, \bibinfo {author} {\bibfnamefont
  {E.}~\bibnamefont {Stern-Taulats}}, \bibinfo {author} {\bibfnamefont
  {A.}~\bibnamefont {Avramenko}}, \bibinfo {author} {\bibfnamefont {N.~D.}\
  \bibnamefont {Mathur}}, \emph {et~al.},\ }\bibfield  {title} {\bibinfo
  {title} {Colossal barocaloric effects near room temperature in plastic
  crystals of neopentylglycol},\ }\href@noop {} {\bibfield  {journal} {\bibinfo
   {journal} {Nature Communications}\ }\textbf {\bibinfo {volume} {10}},\
  \bibinfo {pages} {1} (\bibinfo {year} {2019})}\BibitemShut {NoStop}%
\bibitem [{\citenamefont {Lloveras}\ and\ \citenamefont
  {Tamarit}(2021)}]{lloveras2021advances}%
  \BibitemOpen
  \bibfield  {author} {\bibinfo {author} {\bibfnamefont {P.}~\bibnamefont
  {Lloveras}}\ and\ \bibinfo {author} {\bibfnamefont {J.-L.}\ \bibnamefont
  {Tamarit}},\ }\bibfield  {title} {\bibinfo {title} {Advances and obstacles in
  pressure-driven solid-state cooling: A review of barocaloric materials},\
  }\href@noop {} {\bibfield  {journal} {\bibinfo  {journal} {MRS Energy \&
  Sustainability}\ ,\ \bibinfo {pages} {1}} (\bibinfo {year}
  {2021})}\BibitemShut {NoStop}%
\bibitem [{\citenamefont {Bom}\ \emph {et~al.}(2020)\citenamefont {Bom},
  \citenamefont {Usuda}, \citenamefont {Gigliotti}, \citenamefont {Aguiar},
  \citenamefont {Imamura}, \citenamefont {Paix{\~a}o},\ and\ \citenamefont
  {Carvalho}}]{bom2020tire}%
  \BibitemOpen
  \bibfield  {author} {\bibinfo {author} {\bibfnamefont {N.~M.}\ \bibnamefont
  {Bom}}, \bibinfo {author} {\bibfnamefont {E.~O.}\ \bibnamefont {Usuda}},
  \bibinfo {author} {\bibfnamefont {M.~S.}\ \bibnamefont {Gigliotti}}, \bibinfo
  {author} {\bibfnamefont {D.~J.~M.}\ \bibnamefont {Aguiar}}, \bibinfo {author}
  {\bibfnamefont {W.}~\bibnamefont {Imamura}}, \bibinfo {author} {\bibfnamefont
  {L.~S.}\ \bibnamefont {Paix{\~a}o}},\ and\ \bibinfo {author} {\bibfnamefont
  {A.~M.~G.}\ \bibnamefont {Carvalho}},\ }\bibfield  {title} {\bibinfo {title}
  {Waste tire rubber-based refrigerants for solid-state cooling devices},\
  }\href@noop {} {\bibfield  {journal} {\bibinfo  {journal} {Chinese Journal of
  Polymer Science}\ }\textbf {\bibinfo {volume} {38}},\ \bibinfo {pages} {769 }
  (\bibinfo {year} {2020})}\BibitemShut {NoStop}%
\bibitem [{\citenamefont {Silva}\ \emph {et~al.}(2019)\citenamefont {Silva},
  \citenamefont {Rosado}, \citenamefont {Jasiurkowska-Delaporte}, \citenamefont
  {Silva}, \citenamefont {Piedade}, \citenamefont {Dryzek},\ and\ \citenamefont
  {Eusebio}}]{silva2019ordered}%
  \BibitemOpen
  \bibfield  {author} {\bibinfo {author} {\bibfnamefont {J.~F.}\ \bibnamefont
  {Silva}}, \bibinfo {author} {\bibfnamefont {M.~T.}\ \bibnamefont {Rosado}},
  \bibinfo {author} {\bibfnamefont {M.}~\bibnamefont {Jasiurkowska-Delaporte}},
  \bibinfo {author} {\bibfnamefont {M.~R.}\ \bibnamefont {Silva}}, \bibinfo
  {author} {\bibfnamefont {M.~F.~M.}\ \bibnamefont {Piedade}}, \bibinfo
  {author} {\bibfnamefont {E.}~\bibnamefont {Dryzek}},\ and\ \bibinfo {author}
  {\bibfnamefont {M.~E.~S.}\ \bibnamefont {Eusebio}},\ }\bibfield  {title}
  {\bibinfo {title} {Ordered and plastic crystals in the complex polymorphism
  of pinanediol},\ }\href@noop {} {\bibfield  {journal} {\bibinfo  {journal}
  {Crystal Growth \& Design}\ }\textbf {\bibinfo {volume} {19}},\ \bibinfo
  {pages} {6127} (\bibinfo {year} {2019})}\BibitemShut {NoStop}%
\bibitem [{\citenamefont {Zhu}\ \emph {et~al.}(2020)\citenamefont {Zhu},
  \citenamefont {Chen}, \citenamefont {Zhang}, \citenamefont {Niu},
  \citenamefont {Chen}, \citenamefont {Mo}, \citenamefont {Hu}, \citenamefont
  {Zhang}, \citenamefont {Li}, \citenamefont {Chen} \emph
  {et~al.}}]{zhu2020dissecting}%
  \BibitemOpen
  \bibfield  {author} {\bibinfo {author} {\bibfnamefont {S.}~\bibnamefont
  {Zhu}}, \bibinfo {author} {\bibfnamefont {R.}~\bibnamefont {Chen}}, \bibinfo
  {author} {\bibfnamefont {W.}~\bibnamefont {Zhang}}, \bibinfo {author}
  {\bibfnamefont {X.}~\bibnamefont {Niu}}, \bibinfo {author} {\bibfnamefont
  {W.}~\bibnamefont {Chen}}, \bibinfo {author} {\bibfnamefont {L.}~\bibnamefont
  {Mo}}, \bibinfo {author} {\bibfnamefont {M.}~\bibnamefont {Hu}}, \bibinfo
  {author} {\bibfnamefont {L.}~\bibnamefont {Zhang}}, \bibinfo {author}
  {\bibfnamefont {J.}~\bibnamefont {Li}}, \bibinfo {author} {\bibfnamefont
  {X.}~\bibnamefont {Chen}}, \emph {et~al.},\ }\bibfield  {title} {\bibinfo
  {title} {Dissecting terminal fluorinated regulator of liquid crystals for
  fine-tuning intermolecular interaction and molecular configuration},\
  }\href@noop {} {\bibfield  {journal} {\bibinfo  {journal} {Journal of
  Molecular Liquids}\ }\textbf {\bibinfo {volume} {310}},\ \bibinfo {pages}
  {113225} (\bibinfo {year} {2020})}\BibitemShut {NoStop}%
\bibitem [{\citenamefont {Zhu}\ \emph {et~al.}(2019)\citenamefont {Zhu},
  \citenamefont {MacFarlane}, \citenamefont {Pringle},\ and\ \citenamefont
  {Forsyth}}]{zhu2019organic}%
  \BibitemOpen
  \bibfield  {author} {\bibinfo {author} {\bibfnamefont {H.}~\bibnamefont
  {Zhu}}, \bibinfo {author} {\bibfnamefont {D.~R.}\ \bibnamefont {MacFarlane}},
  \bibinfo {author} {\bibfnamefont {J.~M.}\ \bibnamefont {Pringle}},\ and\
  \bibinfo {author} {\bibfnamefont {M.}~\bibnamefont {Forsyth}},\ }\bibfield
  {title} {\bibinfo {title} {Organic ionic plastic crystals as solid-state
  electrolytes},\ }\href@noop {} {\bibfield  {journal} {\bibinfo  {journal}
  {Trends in Chemistry}\ }\textbf {\bibinfo {volume} {1}},\ \bibinfo {pages}
  {126} (\bibinfo {year} {2019})}\BibitemShut {NoStop}%
\bibitem [{\citenamefont {Chandra}\ \emph {et~al.}(1991)\citenamefont
  {Chandra}, \citenamefont {Ding}, \citenamefont {Lynch},\ and\ \citenamefont
  {Tomilinson}}]{chandra1991transitions}%
  \BibitemOpen
  \bibfield  {author} {\bibinfo {author} {\bibfnamefont {D.}~\bibnamefont
  {Chandra}}, \bibinfo {author} {\bibfnamefont {W.}~\bibnamefont {Ding}},
  \bibinfo {author} {\bibfnamefont {R.~A.}\ \bibnamefont {Lynch}},\ and\
  \bibinfo {author} {\bibfnamefont {J.~J.}\ \bibnamefont {Tomilinson}},\
  }\bibfield  {title} {\bibinfo {title} {Phase transitions in “plastic
  crystals”},\ }\href
  {https://doi.org/https://doi.org/10.1016/0022-5088(91)90042-3} {\bibfield
  {journal} {\bibinfo  {journal} {J. Less-Common Met.}\ }\textbf {\bibinfo
  {volume} {168}},\ \bibinfo {pages} {159 } (\bibinfo {year}
  {1991})}\BibitemShut {NoStop}%
\bibitem [{\citenamefont {Timmermans}(1961)}]{timmermans1961plastic}%
  \BibitemOpen
  \bibfield  {author} {\bibinfo {author} {\bibfnamefont {J.}~\bibnamefont
  {Timmermans}},\ }\bibfield  {title} {\bibinfo {title} {Plastic crystals: a
  historical review},\ }\href@noop {} {\bibfield  {journal} {\bibinfo
  {journal} {Journal of Physics and Chemistry of Solids}\ }\textbf {\bibinfo
  {volume} {18}},\ \bibinfo {pages} {1} (\bibinfo {year} {1961})}\BibitemShut
  {NoStop}%
\bibitem [{\citenamefont {Harada}\ \emph {et~al.}(2018)\citenamefont {Harada},
  \citenamefont {Yoneyama}, \citenamefont {Yokokura}, \citenamefont
  {Takahashi}, \citenamefont {Miura}, \citenamefont {Kitamura},\ and\
  \citenamefont {Inabe}}]{harada2018ferroelectricity}%
  \BibitemOpen
  \bibfield  {author} {\bibinfo {author} {\bibfnamefont {J.}~\bibnamefont
  {Harada}}, \bibinfo {author} {\bibfnamefont {N.}~\bibnamefont {Yoneyama}},
  \bibinfo {author} {\bibfnamefont {S.}~\bibnamefont {Yokokura}}, \bibinfo
  {author} {\bibfnamefont {Y.}~\bibnamefont {Takahashi}}, \bibinfo {author}
  {\bibfnamefont {A.}~\bibnamefont {Miura}}, \bibinfo {author} {\bibfnamefont
  {N.}~\bibnamefont {Kitamura}},\ and\ \bibinfo {author} {\bibfnamefont
  {T.}~\bibnamefont {Inabe}},\ }\bibfield  {title} {\bibinfo {title}
  {Ferroelectricity and piezoelectricity in free-standing polycrystalline films
  of plastic crystals},\ }\href@noop {} {\bibfield  {journal} {\bibinfo
  {journal} {Journal of the American Chemical Society}\ }\textbf {\bibinfo
  {volume} {140}},\ \bibinfo {pages} {346} (\bibinfo {year}
  {2018})}\BibitemShut {NoStop}%
\bibitem [{\citenamefont {Aznar}\ \emph {et~al.}(2020)\citenamefont {Aznar},
  \citenamefont {Lloveras}, \citenamefont {Barrio}, \citenamefont {Negrier},
  \citenamefont {Planes}, \citenamefont {Ma{\~n}osa}, \citenamefont {Mathur},
  \citenamefont {Moya},\ and\ \citenamefont {Tamarit}}]{aznar2020reversible}%
  \BibitemOpen
  \bibfield  {author} {\bibinfo {author} {\bibfnamefont {A.}~\bibnamefont
  {Aznar}}, \bibinfo {author} {\bibfnamefont {P.}~\bibnamefont {Lloveras}},
  \bibinfo {author} {\bibfnamefont {M.}~\bibnamefont {Barrio}}, \bibinfo
  {author} {\bibfnamefont {P.}~\bibnamefont {Negrier}}, \bibinfo {author}
  {\bibfnamefont {A.}~\bibnamefont {Planes}}, \bibinfo {author} {\bibfnamefont
  {L.}~\bibnamefont {Ma{\~n}osa}}, \bibinfo {author} {\bibfnamefont {N.~D.}\
  \bibnamefont {Mathur}}, \bibinfo {author} {\bibfnamefont {X.}~\bibnamefont
  {Moya}},\ and\ \bibinfo {author} {\bibfnamefont {J.-L.}\ \bibnamefont
  {Tamarit}},\ }\bibfield  {title} {\bibinfo {title} {Reversible and
  irreversible colossal barocaloric effects in plastic crystals},\ }\href@noop
  {} {\bibfield  {journal} {\bibinfo  {journal} {Journal of Materials Chemistry
  A}\ } (\bibinfo {year} {2020})}\BibitemShut {NoStop}%
\bibitem [{\citenamefont {Sperling}(2006)}]{sperling-book}%
  \BibitemOpen
  \bibfield  {author} {\bibinfo {author} {\bibfnamefont {L.}~\bibnamefont
  {Sperling}},\ }\href@noop {} {\emph {\bibinfo {title} {Introduction to
  Physical Polymer Science}}},\ \bibinfo {edition} {4th}\ ed.\ (\bibinfo
  {publisher} {Wiley},\ \bibinfo {address} {Hoboken, NJ},\ \bibinfo {year}
  {2006})\BibitemShut {NoStop}%
\bibitem [{\citenamefont {Vijayaram}(2009)}]{vijayaram2009technical}%
  \BibitemOpen
  \bibfield  {author} {\bibinfo {author} {\bibfnamefont {T.~R.}\ \bibnamefont
  {Vijayaram}},\ }\bibfield  {title} {\bibinfo {title} {A technical review on
  rubber},\ }\href@noop {} {\bibfield  {journal} {\bibinfo  {journal} {Int. J.
  Design Manufact. Technol}\ }\textbf {\bibinfo {volume} {3}},\ \bibinfo
  {pages} {25} (\bibinfo {year} {2009})}\BibitemShut {NoStop}%
\bibitem [{\citenamefont {Hosler}\ \emph {et~al.}(1999)\citenamefont {Hosler},
  \citenamefont {Burkett},\ and\ \citenamefont
  {Tarkanian}}]{hosler1988mesoamerica}%
  \BibitemOpen
  \bibfield  {author} {\bibinfo {author} {\bibfnamefont {D.}~\bibnamefont
  {Hosler}}, \bibinfo {author} {\bibfnamefont {S.~L.}\ \bibnamefont
  {Burkett}},\ and\ \bibinfo {author} {\bibfnamefont {M.~J.}\ \bibnamefont
  {Tarkanian}},\ }\bibfield  {title} {\bibinfo {title} {Prehistoric polymers:
  Rubber processing in ancient mesoamerica},\ }\href
  {https://doi.org/10.1126/science.284.5422.1988} {\bibfield  {journal}
  {\bibinfo  {journal} {Science}\ }\textbf {\bibinfo {volume} {284}},\ \bibinfo
  {pages} {1988 } (\bibinfo {year} {1999})}\BibitemShut {NoStop}%
\bibitem [{\citenamefont {Rosato}(2003)}]{rosato2003plastics}%
  \BibitemOpen
  \bibfield  {author} {\bibinfo {author} {\bibfnamefont {D.~V.}\ \bibnamefont
  {Rosato}},\ }\href@noop {} {\emph {\bibinfo {title} {Plastics Engineered
  Product Design}}}\ (\bibinfo  {publisher} {Elsevier},\ \bibinfo {year}
  {2003})\BibitemShut {NoStop}%
\bibitem [{\citenamefont {Furushima}\ \emph {et~al.}(2018)\citenamefont
  {Furushima}, \citenamefont {Schick},\ and\ \citenamefont
  {Toda}}]{furushima2018crystallization}%
  \BibitemOpen
  \bibfield  {author} {\bibinfo {author} {\bibfnamefont {Y.}~\bibnamefont
  {Furushima}}, \bibinfo {author} {\bibfnamefont {C.}~\bibnamefont {Schick}},\
  and\ \bibinfo {author} {\bibfnamefont {A.}~\bibnamefont {Toda}},\ }\bibfield
  {title} {\bibinfo {title} {Crystallization, recrystallization, and melting of
  polymer crystals on heating and cooling examined with fast scanning
  calorimetry},\ }\href@noop {} {\bibfield  {journal} {\bibinfo  {journal}
  {Polymer Crystallization}\ }\textbf {\bibinfo {volume} {1}},\ \bibinfo
  {pages} {e10005} (\bibinfo {year} {2018})}\BibitemShut {NoStop}%
\bibitem [{\citenamefont {Imamura}\ \emph {et~al.}(2020)\citenamefont
  {Imamura}, \citenamefont {Usuda}, \citenamefont {Paix{\~a}o}, \citenamefont
  {Bom}, \citenamefont {Gomes},\ and\ \citenamefont
  {Carvalho}}]{imamurasupergiant}%
  \BibitemOpen
  \bibfield  {author} {\bibinfo {author} {\bibfnamefont {W.}~\bibnamefont
  {Imamura}}, \bibinfo {author} {\bibfnamefont {E.~O.}\ \bibnamefont {Usuda}},
  \bibinfo {author} {\bibfnamefont {L.~S.}\ \bibnamefont {Paix{\~a}o}},
  \bibinfo {author} {\bibfnamefont {N.~M.}\ \bibnamefont {Bom}}, \bibinfo
  {author} {\bibfnamefont {A.~M.}\ \bibnamefont {Gomes}},\ and\ \bibinfo
  {author} {\bibfnamefont {A.~M.~G.}\ \bibnamefont {Carvalho}},\ }\bibfield
  {title} {\bibinfo {title} {Supergiant barocaloric effects in acetoxy silicone
  rubber over a wide temperature range: great potential for solid-state
  cooling},\ }\href@noop {} {\bibfield  {journal} {\bibinfo  {journal} {Chinese
  J. Polym. Sci}\ }\textbf {\bibinfo {volume} {38}},\ \bibinfo {pages} {999 }
  (\bibinfo {year} {2020})}\BibitemShut {NoStop}%
\bibitem [{\citenamefont {Li}\ \emph {et~al.}(2019)\citenamefont {Li},
  \citenamefont {Kawakita}, \citenamefont {Ohira-Kawamura}, \citenamefont
  {Sugahara}, \citenamefont {Wang}, \citenamefont {Wang}, \citenamefont {Chen},
  \citenamefont {Kawaguchi}, \citenamefont {Kawaguchi}, \citenamefont {Ohara}
  \emph {et~al.}}]{li2019colossal}%
  \BibitemOpen
  \bibfield  {author} {\bibinfo {author} {\bibfnamefont {B.}~\bibnamefont
  {Li}}, \bibinfo {author} {\bibfnamefont {Y.}~\bibnamefont {Kawakita}},
  \bibinfo {author} {\bibfnamefont {S.}~\bibnamefont {Ohira-Kawamura}},
  \bibinfo {author} {\bibfnamefont {T.}~\bibnamefont {Sugahara}}, \bibinfo
  {author} {\bibfnamefont {H.}~\bibnamefont {Wang}}, \bibinfo {author}
  {\bibfnamefont {J.}~\bibnamefont {Wang}}, \bibinfo {author} {\bibfnamefont
  {Y.}~\bibnamefont {Chen}}, \bibinfo {author} {\bibfnamefont {S.~I.}\
  \bibnamefont {Kawaguchi}}, \bibinfo {author} {\bibfnamefont {S.}~\bibnamefont
  {Kawaguchi}}, \bibinfo {author} {\bibfnamefont {K.}~\bibnamefont {Ohara}},
  \emph {et~al.},\ }\bibfield  {title} {\bibinfo {title} {Colossal barocaloric
  effects in plastic crystals},\ }\href@noop {} {\bibfield  {journal} {\bibinfo
   {journal} {Nature}\ }\textbf {\bibinfo {volume} {567}},\ \bibinfo {pages}
  {506} (\bibinfo {year} {2019})}\BibitemShut {NoStop}%
\bibitem [{\citenamefont {Usuda}\ \emph {et~al.}(2019)\citenamefont {Usuda},
  \citenamefont {Imamura}, \citenamefont {Bom}, \citenamefont {Paix{\~a}o},\
  and\ \citenamefont {Carvalho}}]{usuda2019giant}%
  \BibitemOpen
  \bibfield  {author} {\bibinfo {author} {\bibfnamefont {E.~O.}\ \bibnamefont
  {Usuda}}, \bibinfo {author} {\bibfnamefont {W.}~\bibnamefont {Imamura}},
  \bibinfo {author} {\bibfnamefont {N.~M.}\ \bibnamefont {Bom}}, \bibinfo
  {author} {\bibfnamefont {L.~S.}\ \bibnamefont {Paix{\~a}o}},\ and\ \bibinfo
  {author} {\bibfnamefont {A.~M.~G.}\ \bibnamefont {Carvalho}},\ }\bibfield
  {title} {\bibinfo {title} {Giant reversible barocaloric effects in nitrile
  butadiene rubber around room temperature},\ }\href@noop {} {\bibfield
  {journal} {\bibinfo  {journal} {ACS Applied Polymer Materials}\ }\textbf
  {\bibinfo {volume} {1}},\ \bibinfo {pages} {1991} (\bibinfo {year}
  {2019})}\BibitemShut {NoStop}%
\bibitem [{\citenamefont {Zhang}\ \emph {et~al.}(2022)\citenamefont {Zhang},
  \citenamefont {Song}, \citenamefont {Qi}, \citenamefont {Zhang},
  \citenamefont {Zhang}, \citenamefont {Yu}, \citenamefont {Li}, \citenamefont
  {Zhang},\ and\ \citenamefont {Li}}]{zhang2022colossal}%
  \BibitemOpen
  \bibfield  {author} {\bibinfo {author} {\bibfnamefont {K.}~\bibnamefont
  {Zhang}}, \bibinfo {author} {\bibfnamefont {R.}~\bibnamefont {Song}},
  \bibinfo {author} {\bibfnamefont {J.}~\bibnamefont {Qi}}, \bibinfo {author}
  {\bibfnamefont {Z.}~\bibnamefont {Zhang}}, \bibinfo {author} {\bibfnamefont
  {Z.}~\bibnamefont {Zhang}}, \bibinfo {author} {\bibfnamefont
  {C.}~\bibnamefont {Yu}}, \bibinfo {author} {\bibfnamefont {K.}~\bibnamefont
  {Li}}, \bibinfo {author} {\bibfnamefont {Z.}~\bibnamefont {Zhang}},\ and\
  \bibinfo {author} {\bibfnamefont {B.}~\bibnamefont {Li}},\ }\bibfield
  {title} {\bibinfo {title} {Colossal barocaloric effect in carboranes as a
  performance tradeoff},\ }\href@noop {} {\bibfield  {journal} {\bibinfo
  {journal} {Advanced Functional Materials}\ ,\ \bibinfo {pages} {2112622}}
  (\bibinfo {year} {2022})}\BibitemShut {NoStop}%
\bibitem [{\citenamefont {McLinden}(2005)}]{mclinden2005thermophysical}%
  \BibitemOpen
  \bibfield  {author} {\bibinfo {author} {\bibfnamefont {M.~O.}\ \bibnamefont
  {McLinden}},\ }\bibfield  {title} {\bibinfo {title} {Thermophysical
  properties of refrigerants, ashrae handbook--fundamentals (2005 revision)},\
  }\href@noop {} {\  (\bibinfo {year} {2005})}\BibitemShut {NoStop}%
\bibitem [{\citenamefont {Aznar}\ \emph {et~al.}(2021)\citenamefont {Aznar},
  \citenamefont {Negrier}, \citenamefont {Planes}, \citenamefont {Ma{\~n}osa},
  \citenamefont {Stern-Taulats}, \citenamefont {Moya}, \citenamefont {Barrio},
  \citenamefont {Tamarit},\ and\ \citenamefont
  {Lloveras}}]{aznar2021reversible}%
  \BibitemOpen
  \bibfield  {author} {\bibinfo {author} {\bibfnamefont {A.}~\bibnamefont
  {Aznar}}, \bibinfo {author} {\bibfnamefont {P.}~\bibnamefont {Negrier}},
  \bibinfo {author} {\bibfnamefont {A.}~\bibnamefont {Planes}}, \bibinfo
  {author} {\bibfnamefont {L.}~\bibnamefont {Ma{\~n}osa}}, \bibinfo {author}
  {\bibfnamefont {E.}~\bibnamefont {Stern-Taulats}}, \bibinfo {author}
  {\bibfnamefont {X.}~\bibnamefont {Moya}}, \bibinfo {author} {\bibfnamefont
  {M.}~\bibnamefont {Barrio}}, \bibinfo {author} {\bibfnamefont {J.-L.}\
  \bibnamefont {Tamarit}},\ and\ \bibinfo {author} {\bibfnamefont
  {P.}~\bibnamefont {Lloveras}},\ }\bibfield  {title} {\bibinfo {title}
  {Reversible colossal barocaloric effects near room temperature in
  1-{X}-adamantane ({X}= {C}l, {B}r) plastic crystals},\ }\href@noop {}
  {\bibfield  {journal} {\bibinfo  {journal} {Applied Materials Today}\
  }\textbf {\bibinfo {volume} {23}},\ \bibinfo {pages} {101023} (\bibinfo
  {year} {2021})}\BibitemShut {NoStop}%
\bibitem [{\citenamefont {Bazyleva}\ \emph {et~al.}(2005)\citenamefont
  {Bazyleva}, \citenamefont {Blokhin}, \citenamefont {Kabo}, \citenamefont
  {Kabo},\ and\ \citenamefont {Paulechka}}]{bazyleva2005thermodynamic}%
  \BibitemOpen
  \bibfield  {author} {\bibinfo {author} {\bibfnamefont {A.~B.}\ \bibnamefont
  {Bazyleva}}, \bibinfo {author} {\bibfnamefont {A.~V.}\ \bibnamefont
  {Blokhin}}, \bibinfo {author} {\bibfnamefont {G.~J.}\ \bibnamefont {Kabo}},
  \bibinfo {author} {\bibfnamefont {A.~G.}\ \bibnamefont {Kabo}},\ and\
  \bibinfo {author} {\bibfnamefont {Y.~U.}\ \bibnamefont {Paulechka}},\
  }\bibfield  {title} {\bibinfo {title} {Thermodynamic properties of
  1-bromoadamantane in the condensed state and molecular disorder in its
  crystals},\ }\href@noop {} {\bibfield  {journal} {\bibinfo  {journal} {The
  Journal of Chemical Thermodynamics}\ }\textbf {\bibinfo {volume} {37}},\
  \bibinfo {pages} {643} (\bibinfo {year} {2005})}\BibitemShut {NoStop}%
\bibitem [{\citenamefont {Carvalho}\ \emph {et~al.}(2018)\citenamefont
  {Carvalho}, \citenamefont {Imamura}, \citenamefont {Usuda},\ and\
  \citenamefont {Bom}}]{carvalho2018giant}%
  \BibitemOpen
  \bibfield  {author} {\bibinfo {author} {\bibfnamefont {A.~M.~G.}\
  \bibnamefont {Carvalho}}, \bibinfo {author} {\bibfnamefont {W.}~\bibnamefont
  {Imamura}}, \bibinfo {author} {\bibfnamefont {E.~O.}\ \bibnamefont {Usuda}},\
  and\ \bibinfo {author} {\bibfnamefont {N.~M.}\ \bibnamefont {Bom}},\
  }\bibfield  {title} {\bibinfo {title} {Giant room-temperature barocaloric
  effects in {PDMS} rubber at low pressures},\ }\href@noop {} {\bibfield
  {journal} {\bibinfo  {journal} {European Polymer Journal}\ }\textbf {\bibinfo
  {volume} {99}},\ \bibinfo {pages} {212} (\bibinfo {year} {2018})}\BibitemShut
  {NoStop}%
\bibitem [{\citenamefont {Bom}\ \emph {et~al.}(2018)\citenamefont {Bom},
  \citenamefont {Imamura}, \citenamefont {Usuda}, \citenamefont {Paix{\~a}o},\
  and\ \citenamefont {Carvalho}}]{bom2018giant}%
  \BibitemOpen
  \bibfield  {author} {\bibinfo {author} {\bibfnamefont {N.}~\bibnamefont
  {Bom}}, \bibinfo {author} {\bibfnamefont {W.}~\bibnamefont {Imamura}},
  \bibinfo {author} {\bibfnamefont {E.}~\bibnamefont {Usuda}}, \bibinfo
  {author} {\bibfnamefont {L.}~\bibnamefont {Paix{\~a}o}},\ and\ \bibinfo
  {author} {\bibfnamefont {A.}~\bibnamefont {Carvalho}},\ }\bibfield  {title}
  {\bibinfo {title} {Giant barocaloric effects in natural rubber: {A} relevant
  step toward solid-state cooling},\ }\href@noop {} {\bibfield  {journal}
  {\bibinfo  {journal} {ACS Macro Letters}\ }\textbf {\bibinfo {volume} {7}},\
  \bibinfo {pages} {31} (\bibinfo {year} {2018})}\BibitemShut {NoStop}%
\bibitem [{\citenamefont {Bocca}\ \emph {et~al.}(2021)\citenamefont {Bocca},
  \citenamefont {Favaro}, \citenamefont {Alves}, \citenamefont {Carvalho},
  \citenamefont {Barbosa~Jr}, \citenamefont {dos Santos}, \citenamefont
  {Colman}, \citenamefont {Concei{\c{c}}{\~a}o}, \citenamefont {Caglioni},\
  and\ \citenamefont {Radovanovic}}]{bocca2021giant}%
  \BibitemOpen
  \bibfield  {author} {\bibinfo {author} {\bibfnamefont {J.~R.}\ \bibnamefont
  {Bocca}}, \bibinfo {author} {\bibfnamefont {S.~L.}\ \bibnamefont {Favaro}},
  \bibinfo {author} {\bibfnamefont {C.~S.}\ \bibnamefont {Alves}}, \bibinfo
  {author} {\bibfnamefont {A.~M.}\ \bibnamefont {Carvalho}}, \bibinfo {author}
  {\bibfnamefont {J.~R.}\ \bibnamefont {Barbosa~Jr}}, \bibinfo {author}
  {\bibfnamefont {A.}~\bibnamefont {dos Santos}}, \bibinfo {author}
  {\bibfnamefont {F.~C.}\ \bibnamefont {Colman}}, \bibinfo {author}
  {\bibfnamefont {W.~A. d.~S.}\ \bibnamefont {Concei{\c{c}}{\~a}o}}, \bibinfo
  {author} {\bibfnamefont {C.}~\bibnamefont {Caglioni}},\ and\ \bibinfo
  {author} {\bibfnamefont {E.}~\bibnamefont {Radovanovic}},\ }\bibfield
  {title} {\bibinfo {title} {Giant barocaloric effect in commercial
  polyurethane},\ }\href@noop {} {\bibfield  {journal} {\bibinfo  {journal}
  {Polymer Testing}\ }\textbf {\bibinfo {volume} {100}},\ \bibinfo {pages}
  {107251} (\bibinfo {year} {2021})}\BibitemShut {NoStop}%
\bibitem [{\citenamefont {Jenau}\ \emph {et~al.}(1996)\citenamefont {Jenau},
  \citenamefont {Reuter}, \citenamefont {Tamarit},\ and\ \citenamefont
  {W{\"u}rflinger}}]{jenau1996crystal}%
  \BibitemOpen
  \bibfield  {author} {\bibinfo {author} {\bibfnamefont {M.}~\bibnamefont
  {Jenau}}, \bibinfo {author} {\bibfnamefont {J.}~\bibnamefont {Reuter}},
  \bibinfo {author} {\bibfnamefont {J.~L.}\ \bibnamefont {Tamarit}},\ and\
  \bibinfo {author} {\bibfnamefont {A.}~\bibnamefont {W{\"u}rflinger}},\
  }\bibfield  {title} {\bibinfo {title} {Crystal and p{VT} data and
  thermodynamics of the phase transitions of 2-methyl-2-nitropropane},\
  }\href@noop {} {\bibfield  {journal} {\bibinfo  {journal} {Journal of the
  Chemical Society, Faraday Transactions}\ }\textbf {\bibinfo {volume} {92}},\
  \bibinfo {pages} {1899} (\bibinfo {year} {1996})}\BibitemShut {NoStop}%
\bibitem [{\citenamefont {Salud}\ \emph {et~al.}(1999)\citenamefont {Salud},
  \citenamefont {L{\'o}pez}, \citenamefont {Barrio},\ and\ \citenamefont
  {Tamarit}}]{salud1999two}%
  \BibitemOpen
  \bibfield  {author} {\bibinfo {author} {\bibfnamefont {J.}~\bibnamefont
  {Salud}}, \bibinfo {author} {\bibfnamefont {D.~O.}\ \bibnamefont
  {L{\'o}pez}}, \bibinfo {author} {\bibfnamefont {M.}~\bibnamefont {Barrio}},\
  and\ \bibinfo {author} {\bibfnamefont {J.~L.}\ \bibnamefont {Tamarit}},\
  }\bibfield  {title} {\bibinfo {title} {Two-component systems of isomorphous
  orientationally disordered crystals. part 1 packing of the mixed crystals},\
  }\href@noop {} {\bibfield  {journal} {\bibinfo  {journal} {Journal of
  Materials Chemistry}\ }\textbf {\bibinfo {volume} {9}},\ \bibinfo {pages}
  {909} (\bibinfo {year} {1999})}\BibitemShut {NoStop}%
\bibitem [{\citenamefont {Dart}\ \emph {et~al.}(1942)\citenamefont {Dart},
  \citenamefont {Anthony},\ and\ \citenamefont {Guth}}]{dart1942rise}%
  \BibitemOpen
  \bibfield  {author} {\bibinfo {author} {\bibfnamefont {S.}~\bibnamefont
  {Dart}}, \bibinfo {author} {\bibfnamefont {R.}~\bibnamefont {Anthony}},\ and\
  \bibinfo {author} {\bibfnamefont {E.}~\bibnamefont {Guth}},\ }\bibfield
  {title} {\bibinfo {title} {Rise of temperature on fast stretching of
  synthetics and natural rubbers},\ }\href@noop {} {\bibfield  {journal}
  {\bibinfo  {journal} {Industrial \& Engineering Chemistry}\ }\textbf
  {\bibinfo {volume} {34}},\ \bibinfo {pages} {1340} (\bibinfo {year}
  {1942})}\BibitemShut {NoStop}%
\bibitem [{\citenamefont {Bom}\ \emph {et~al.}(2017)\citenamefont {Bom},
  \citenamefont {Usuda}, \citenamefont {Guimar{\~a}es}, \citenamefont
  {Coelho},\ and\ \citenamefont {Carvalho}}]{bom2017note}%
  \BibitemOpen
  \bibfield  {author} {\bibinfo {author} {\bibfnamefont {N.~M.}\ \bibnamefont
  {Bom}}, \bibinfo {author} {\bibfnamefont {E.~O.}\ \bibnamefont {Usuda}},
  \bibinfo {author} {\bibfnamefont {G.~M.}\ \bibnamefont {Guimar{\~a}es}},
  \bibinfo {author} {\bibfnamefont {A.~A.}\ \bibnamefont {Coelho}},\ and\
  \bibinfo {author} {\bibfnamefont {A.~M.~G.}\ \bibnamefont {Carvalho}},\
  }\bibfield  {title} {\bibinfo {title} {Note: {E}xperimental setup for
  measuring the barocaloric effect in polymers: {A}pplication to natural
  rubber},\ }\href@noop {} {\bibfield  {journal} {\bibinfo  {journal} {Review
  of Scientific Instruments}\ }\textbf {\bibinfo {volume} {88}},\ \bibinfo
  {pages} {046103} (\bibinfo {year} {2017})}\BibitemShut {NoStop}%
\bibitem [{\citenamefont {Toki}\ \emph {et~al.}(2000)\citenamefont {Toki},
  \citenamefont {Fujimaki},\ and\ \citenamefont {Okuyama}}]{toki2000strain}%
  \BibitemOpen
  \bibfield  {author} {\bibinfo {author} {\bibfnamefont {S.}~\bibnamefont
  {Toki}}, \bibinfo {author} {\bibfnamefont {T.}~\bibnamefont {Fujimaki}},\
  and\ \bibinfo {author} {\bibfnamefont {M.}~\bibnamefont {Okuyama}},\
  }\bibfield  {title} {\bibinfo {title} {Strain-induced crystallization of
  natural rubber as detected real-time by wide-angle x-ray diffraction
  technique},\ }\href@noop {} {\bibfield  {journal} {\bibinfo  {journal}
  {Polymer}\ }\textbf {\bibinfo {volume} {41}},\ \bibinfo {pages} {5423}
  (\bibinfo {year} {2000})}\BibitemShut {NoStop}%
\bibitem [{\citenamefont {Le~Cam}(2017)}]{le2017energy}%
  \BibitemOpen
  \bibfield  {author} {\bibinfo {author} {\bibfnamefont {J.-B.}\ \bibnamefont
  {Le~Cam}},\ }\bibfield  {title} {\bibinfo {title} {Energy storage due to
  strain-induced crystallization in natural rubber: the physical origin of the
  mechanical hysteresis},\ }\href@noop {} {\bibfield  {journal} {\bibinfo
  {journal} {Polymer}\ }\textbf {\bibinfo {volume} {127}},\ \bibinfo {pages}
  {166} (\bibinfo {year} {2017})}\BibitemShut {NoStop}%
\end{thebibliography}%
\end{document}